\documentclass[useAMS,usenatbib]{mn2e}

\usepackage{amssymb}
\usepackage{amsfonts}
\usepackage{mncite}
\usepackage{epsfig}
\usepackage{psfig}

\newcommand{\de}{\mbox{d}}

\defcitealias{scheuer96}{SF}

\begin{document}

\title[Evolution of misaligned discs and black holes]{The evolution of
misaligned accretion discs and spinning black holes}

\author[G. Lodato and J. E. Pringle]{G. Lodato and J. E. Pringle\\
 Institute of Astronomy, Madingley Road, Cambridge, CB3 0HA}

\maketitle

\begin{abstract}

In this paper we consider the process of alignment of a spinning black
hole and a surrounding misaligned accretion disc. We use a simplified
set of equations, that describe the evolution of the system in the case
where the propagation of warping disturbances in the accretion disc
occurs diffusively, a situation likely to be common in the thin discs
in Active Galactic Nuclei (AGN). We also allow the direction of the
hole spin to move under the action of the disc torques. In such a way,
the evolution of the hole-disc system is computed self-consistently. We
consider a number of different situations and we explore the relevant
parameter range, by varying the location of the warp radius $R_{\rm w}$
and the propagation speed of the warp. We find that the dissipation
associated with the twisting of the disc results in a large increase in
the accretion rate through the disc, so that AGN accreting from a
misaligned disc are likely to be significantly more luminous than those
accreting from a flat disc. We compute explicitly the time-scales for
the warping of the disc and for the alignment process and compare our
results with earlier estimates based on simplified steady-state
solutions.  We also confirm earlier predictions that, under appropriate
circumstances, accretion can proceed in a counter-aligned fashion, so
that the accreted material will spin-down the hole, rather than spinning
it up. Our results have implication in a number of different
observational features of AGN such as the orientation and shape of
jets, the shape of X-ray iron lines, and the possibility of obscuration
and absorption of X-ray by the outer disc as well as the general issue
of the spin history of black holes during their growth.

\end{abstract}


\section{Introduction}

There are strong observational and theoretical grounds for believing
that accretion discs around black holes may be twisted or warped, in
the sense that the plane of the Keplerian orbits varies continuously
with radius and with time. From the observational point of view, warped
discs have been observed in the nuclei of active galaxies (AGN) at
relatively large distances from the central supermassive black hole
($\sim 1$ pc), especially though water maser observation (for example:
in NGC4258, \citealt{herrnstein96}, or in Circinus,
\citealt{greenhill03}). A warped inner accretion disc might also be the
explanation of the apparent lack of correlation between the direction
of the radio jets emanating from AGN and the plane of the host galaxy
disc \citep{kinney00,schmitt02}. Modelling of accretion events onto the
cores of galaxies seems to indicate that they occur at random
orientations \citep{kendall03,saitoh04}. In addition, a warp in the
disc might lead to interesting phenomena of obscuration of the central
engine in an AGN \citep{phinney89,greenhill03,nayak05}.

From a theoretical perspective, there are a number of possible physical
mechanisms that can cause the disc to be warped. On the large scale, a
warp might be produced through the interaction of the disc with a
smaller mass companion orbiting in an inclined orbit. However, closer
to the black hole, at distances of the order of $\approx 10^2-10^3
R_{\rm S}$, where $R_{\rm S}$ is the Schwarzschild radius of the black
hole, the most likely cause of warping is due to relativistic effects.
In particular, if the black hole is spinning and the spin axis of the
hole is misaligned with the axis of rotation of the disc (or,
equivalently, if the hole's angular momentum is misaligned with respect
to the angular momentum of the accreting matter), then the
Lense-Thirring precession produces a warp in the disc. This process is
called the Bardeen-Petterson effect \citep{bardeen75}.  Since the rate
of the Lense-Thirring precession is a strong function of radius
$\propto R^{-3}$, the general result is that the inner parts of the
disc tends to align the direction (but not necessarily the sense, see
below) of its specific angular momentum with that of the hole, while
outside a typical radius, the warp radius, $R_{\rm w}\approx 10^2
R_{\rm S}$ \citep{natarajan98,king05}, the disc tends to retain its
original direction of rotation.

In this paper, we explore the time-dependent evolution of the process
of alignment between an accretion disc and a spinning black hole. We
consider the case where the propagation of the warp occurs
diffusively. This occurs \citep{pappringle83} when the dimensionless
viscosity coefficient in the disc (here called $\alpha_1$, see below)
is larger than the disc aspect ratio $H/R$ (which should be the case
for typical AGN discs, \citealt{wijers99}). Here we are not concerned
with the detailed physical mechanism responsible for the propagation of
the warp, the discussion of which would require large 3D numerical
simulations (c.f. \citealt{nelson00,fragile05}). Rather, we explore
here the long term evolution of the system by using a simplified 1D
approach \citep{pringle92}.  Additionally, in our implementation, we
allow the black hole spin direction to evolve as a consequence of the
mutual torques with the hole. In this way, the evolution of the hole
spin and the warping of the disc, which might occur on similar
time-scales, are computed simultaneously and self-consistently. We
explicitly evaluate the above-mentioned time-scales, and we also
determine the effect that the warping process has on the accretion
phenomenon. In particular, we find that the additional dissipation
associated with the twisting of the disc leads to a large enhancement
in the accretion rate (up to a factor $\sim 10$).

We also address the issue of whether the disc and the hole angular
momenta might end up being counter-aligned. This possibility appeared
ruled out by the analysis of \citet{scheuer96} (hereafter SF).
However, \citet{king05} have recently challenged this conclusion, and,
using simple geometrical arguments, find that counter-alignment should
take place when the magnitude of the hole angular momentum $J_{\rm h}$,
the disc angular momentum $J_{\rm d}$ and their initial misalignment
are in the following relation:

\begin{equation}
\label{eq:king}
\frac{J_{\rm d}}{J_{\rm h}}<-2\cos\theta,
\end{equation}
where $\theta$ is the initial angle between the two angular momenta.
\citet{king05} argue that \citetalias{scheuer96} did not find
counter-alignment because they implicitly assume that $J_{\rm d}$ was
infinitely large, therefore making inequality (\ref{eq:king})
impossible to be satisfied. Here, we re-examine this issue by
performing a number of simulations with given $J_{\rm d}$ and $J_{\rm
h}$ and conclude that indeed counter-alignment is possible when
inequality (\ref{eq:king}) is satisfied.  When counter-alignment
occurs, accretion of matter onto the hole causes the hole to spin-down,
rather than spin-up. The possibility of counter-alignment is therefore
particularly relevant to the description of the spin history of black
holes \citep{volonteri05} and to the determination of the innermost
stable orbit, which in turn determines the efficiency of conversion of
matter into luminosity.

The paper is organized as follows. In Section 2 we describe the basic
set of equations that we use to determine the evolution of the accretion
disc and of the spinning black hole. In Section 3 we examine more
thoroughly the steady-state equations found by \citetalias{scheuer96},
discussing their applicability and their limitations. In Section 4 we
perform a number of simulations describing the evolution of ``warped
spreading rings'', where we consider the evolution of thin rings of
matter, initially inclined with respect to the hole spin direction, as
they are accreted and twisted by the black hole. In Section 5 we
discuss our results and draw our conclusions 

\section{Basic equations}

In this Section we briefly describe the model that we use to follow the
evolution of the system. We consider a system comprising a black hole
at $R=0$ (note that here $R$ is a purely spherical coordinate and not
the cylindrical radius), with spin angular momentum ${\bf J}_{\rm h}$,
surrounded by a disc, with surface density $\Sigma$ and specific
angular momentum density ${\bf L}(R)$, in Keplerian rotation around the
black hole. Since we expect the interesting disc dynamics to occur in
the region of the warp radius $R_{\rm w}\gg R_{\rm S}$, treating the
black hole as a point mass and using Newtonian dynamics is adequate for
our present purposes. The modulus of the specific angular momentum
density is thus given by $L(R)=\Sigma\sqrt{GMR}$, where $M$ is the mass
of the black hole. The unit vector indicating the local direction of
the specific angular momentum in the disc is denoted by ${\bf
l}(R)$. We further denote by ${\bf J}_{\rm d}$ the total angular
momentum of the disc:

\begin{equation}
{\bf J}_{\rm d}=\int 2\pi R {\bf L}(R) \de R
\end{equation}

We follow the formalism of \citet{pringle92} and introduce the two
viscosities $\nu_1$ and $\nu_2$, where $\nu_1$ is the standard shear
viscosity in a flat disc, and $\nu_2$ is the viscosity associated with
vertical shear motions and which describes the diffusion of warping
disturbances through the disc. The equation of motion for the specific
angular momentum in the disc is \citep{pringle92}:

\begin{eqnarray}
\label{eq:pringle}
\nonumber\frac{\partial{\bf L}}{\partial
t} = &&\frac{3}{R}\frac{\partial}{\partial R}
\left[\frac{R^{1/2}}{\Sigma}\frac{\partial}{\partial R} 
(\nu_1\Sigma R^{1/2}){\bf L}\right]\\
&+&\frac{1}{R}\frac{\partial}{\partial R}\left[\left(\nu_2 R^2\left| 
\frac{\partial{\bf l}}{\partial R}\right|^2-\frac{3}{2}\nu_1\right)
{\bf L}\right]\\
\nonumber&+&\frac{1}{R}\frac{\partial}{\partial
R}\left(\frac{1}{2}\nu_2R|{\bf L}|\frac{\partial{\bf l}}{\partial R}
\right).
\end{eqnarray}

Note that the above equation is a generalization of the standard
accretion disc equation (e.g. \citealt{pringle81}) and reduces to it in
the case of a flat disc ($\partial{\bf l}/\partial R=0$). The viscosity
coefficients $\nu_1$ and $\nu_2$, of course, are in general a function
of local disc properites as well as of $R$. However, in the rest of the
paper, we will explore the simple case where both $\nu_1$ and $\nu_2$
are constants, independent on $R$. \citet{pappringle83} have shown that
in the linear approximation corresponding to small warp angles they are
related through $\nu_2/\nu_1=1/2\alpha_1^2$, where $\alpha_1$ is the
standard $\alpha$ parameter associated with the viscosity $\nu_1$
\citep{shakura73}. In this picture, therefore, the stress associated
with warp propagation is much larger than the viscous stress, since
$\alpha_1$ is generally much smaller than unity. However, the actual
value of $\nu_2$ in the non-linear case has still to be determined (see
discussion in Section 2.1. below).

Equation (\ref{eq:pringle}) describes the evolution of a warped disc
subject only to internal torques. In order to study the
Bardeen-Petterson effect we add a term on the right-hand side of
equation (\ref{eq:pringle}), that describes the relative torque between
the disc and the hole due to Lense-Thirring precession:

\begin{equation}
\left.\frac{\partial{\bf L}}{\partial t}\right|_{\rm
LT}={\bf\Omega}_{\rm p}\times{\bf L},
\label{eq:lense}
\end{equation}
where the precession rate ${\bf\Omega}_{\rm p}$ is given by (e.g.
\citealt{kumar85}) ${\bf\Omega}_{\rm p}=\mbox{\boldmath$\omega$}_{\rm
p}/R^3$, with:

\begin{equation}
\mbox{\boldmath$\omega$}_{\rm p}=\frac{2G{\bf J}_{\rm h}}{c^2},
\end{equation}
where $c$ is the speed of light. This is valid for $R\gg R_{\rm S}$. 

In most of our simulations, we also allow the black hole spin to
evolve as a consequence of the torques exerted by the disc. The
equation of motion for the hole spin follows directly from Newton's
third law and we have:

\begin{equation}
\frac{\de{\bf J}_{\rm h}}{\de t}=-2\pi\int{\bf\Omega}_{\rm p}\times{\bf
L} R\de R,
\label{eq:hole}
\end{equation}
where the integral is taken over the disc surface. Note that, since the
torque on the black hole is perpendicular to the hole angular momentum,
the modulus of ${\bf J}_{\rm h}$ is constant with time. The only
effects of the relative torques are to make the hole spin change
direction or precess and affect alignment with the angular momentum of
the disc.  The calculations here neglect the changes in the magnitude
of the spin of the hole caused by the accretion of matter onto it. In
general we expect the accretion induced change in the hole spin to
occur on a longer time-scale than the disc induced change in spin
direction both because of the lever-arm effect at $R_{\rm w}$
\citep{rees78,natarajan98} and because warp propagation ($\nu_2$)
typically occurs faster than mass propagation ($\nu_1$). We discuss
this issue more in detail below.

We solve equations (\ref{eq:pringle}) and (\ref{eq:hole}) numerically
using the scheme described in \citet{pringle92}. We employ a
logarithmically spaced radial grid spanning three to four orders of
magnitude for most simulations. Equation (\ref{eq:hole}) is solved
using a standard fourth-order Runge-Kutta method. 

\begin{figure*}
\centerline{\epsfig{figure=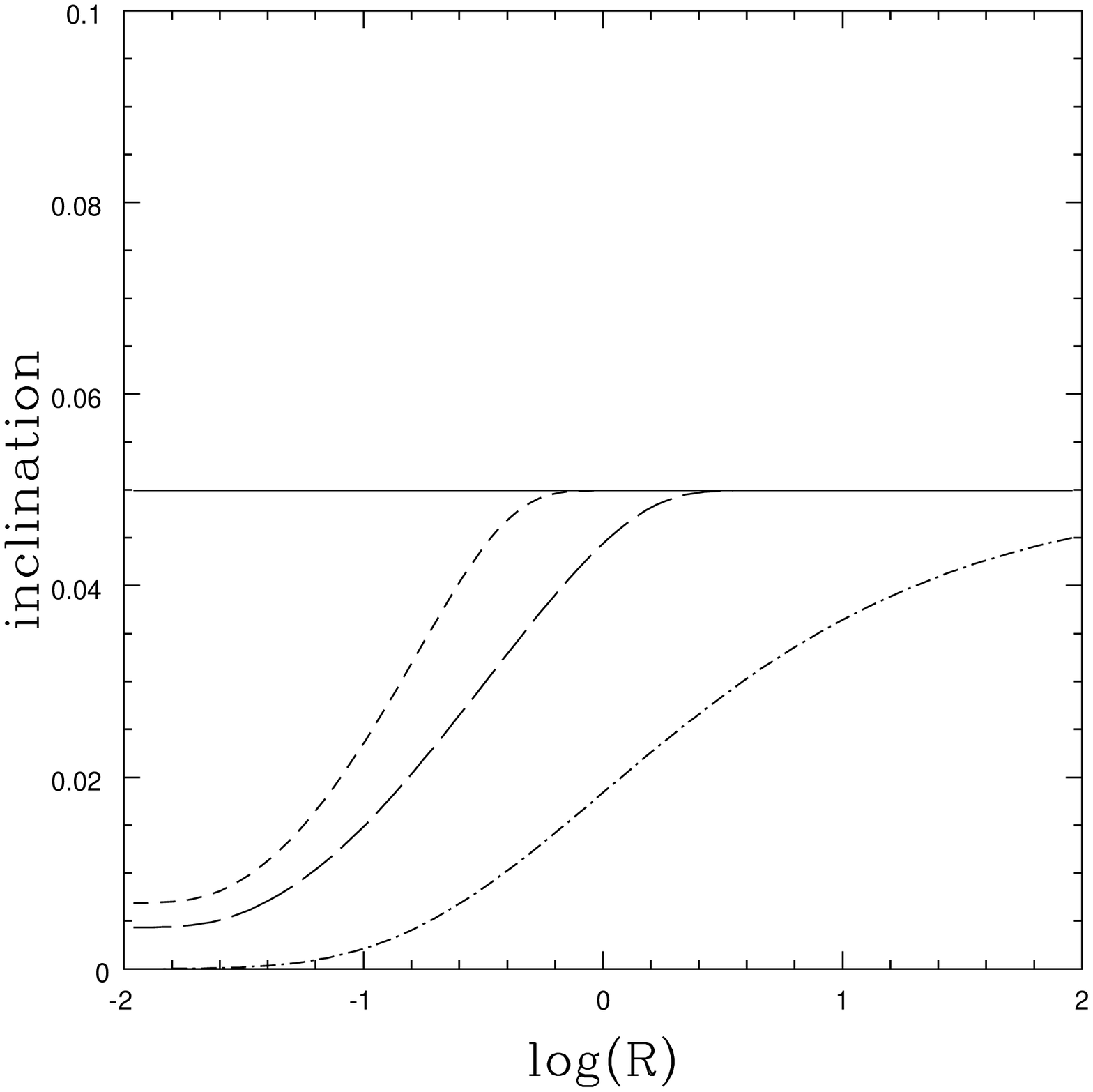,width=0.5\textwidth}
            \epsfig{figure=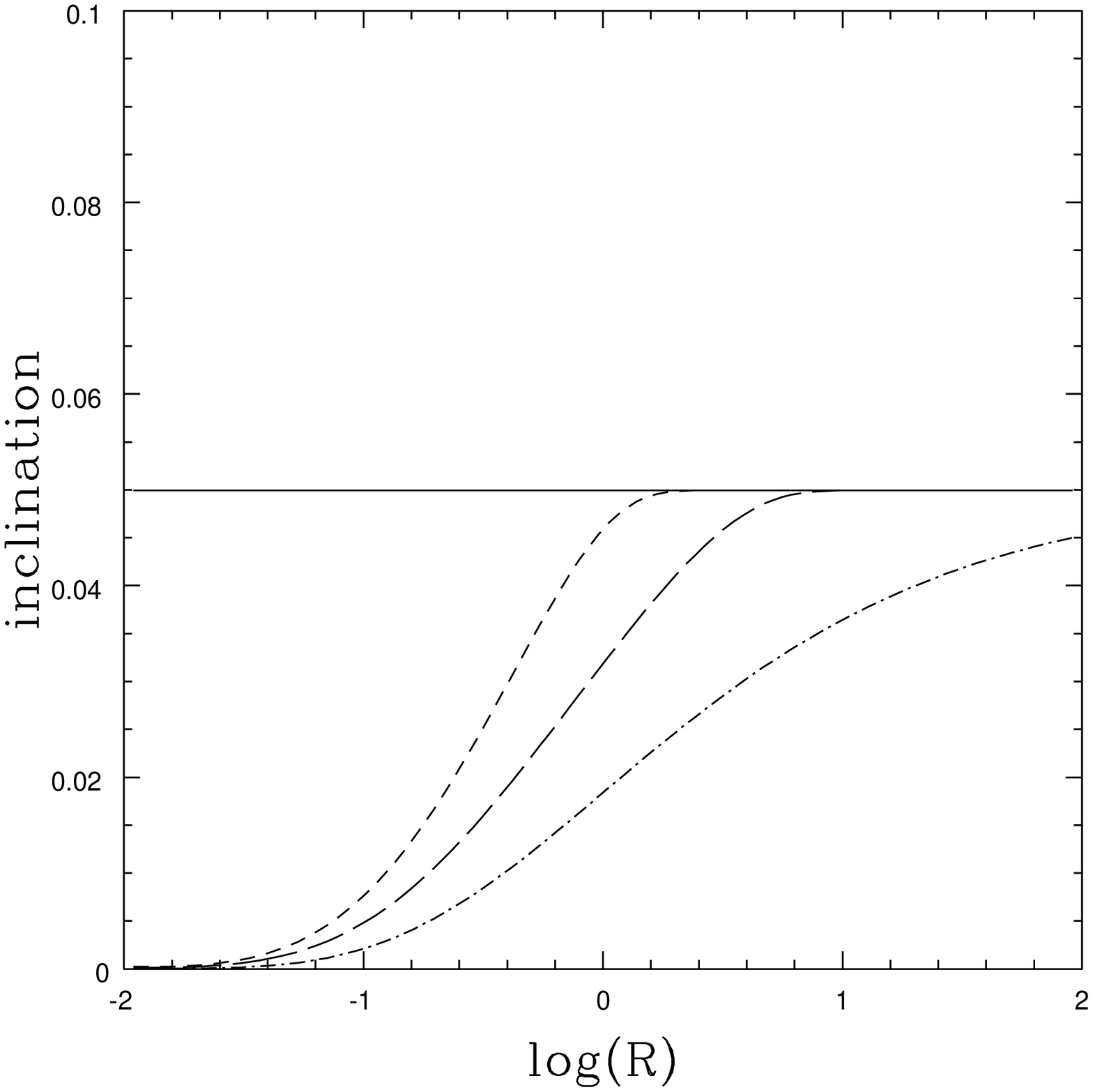,width=0.5\textwidth}}
\caption{Evolution of the warp in the case where $R_{\rm w}=\omega_{\rm
p}/\nu_1=0.25$ ($\log R_{\rm w}=-0.6$) and $\nu_2/\nu_1=1$ (left panel)
and $\nu_2/\nu_1=10$ (right panel). The lines refer to $t=0$ (solid
line), $t=0.05$ (short-dashed line) and $t=1$ (long-dashed line), where
the times are expressed in units of the viscous time ($R^2/\nu_1$) at
$R=1$. The dot-dashed line shows the \citetalias{scheuer96} solution in
this case. At time $t$, the warped portion of the disc extends out to a
radius such that $t\approx R^2/\nu_2$, showing that the warp indeed
diffuses out at this time-scale.}
\label{fig:large}
\end{figure*}

\begin{figure*}
\centerline{\epsfig{figure=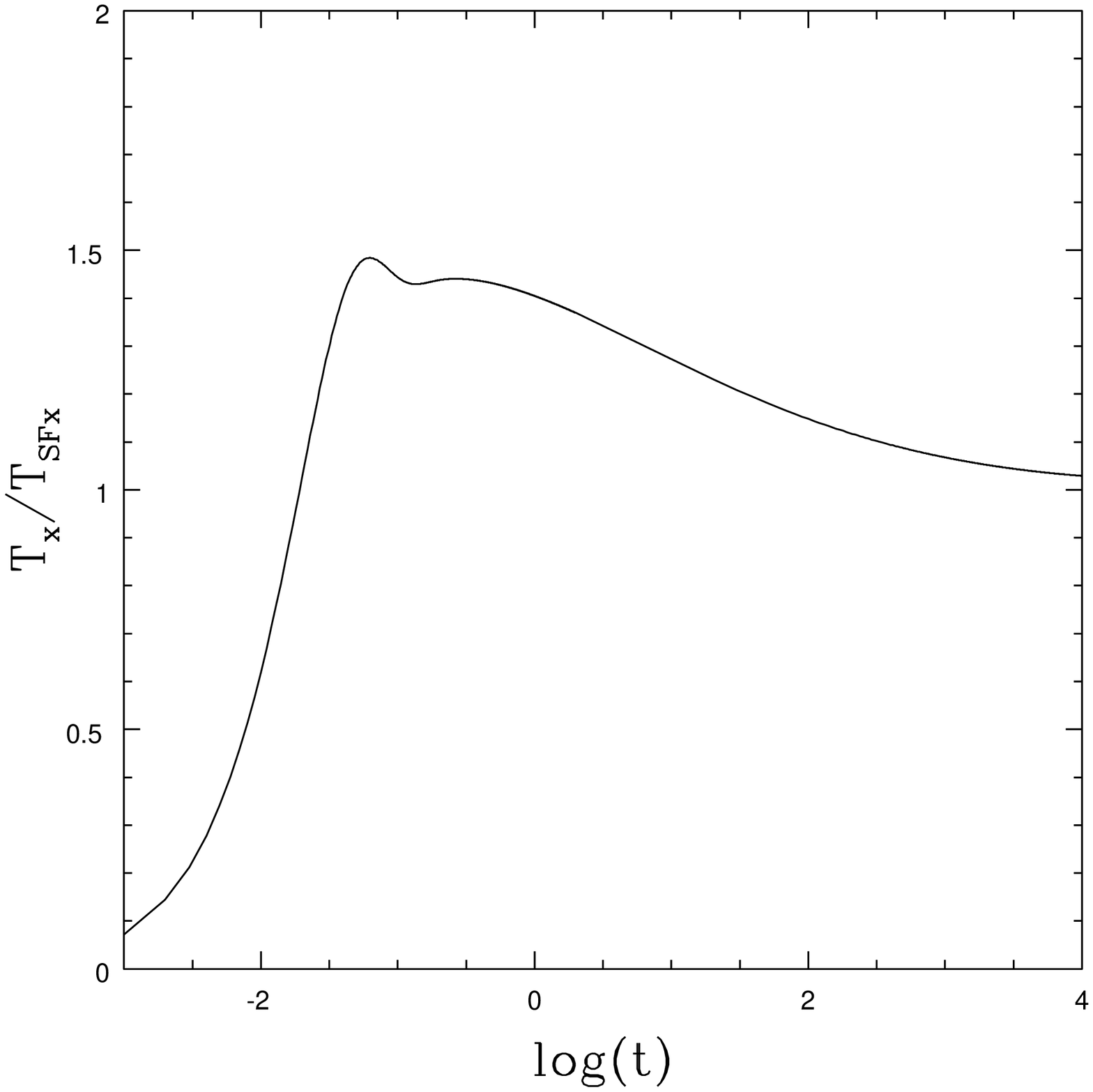,width=0.5\textwidth}
            \epsfig{figure=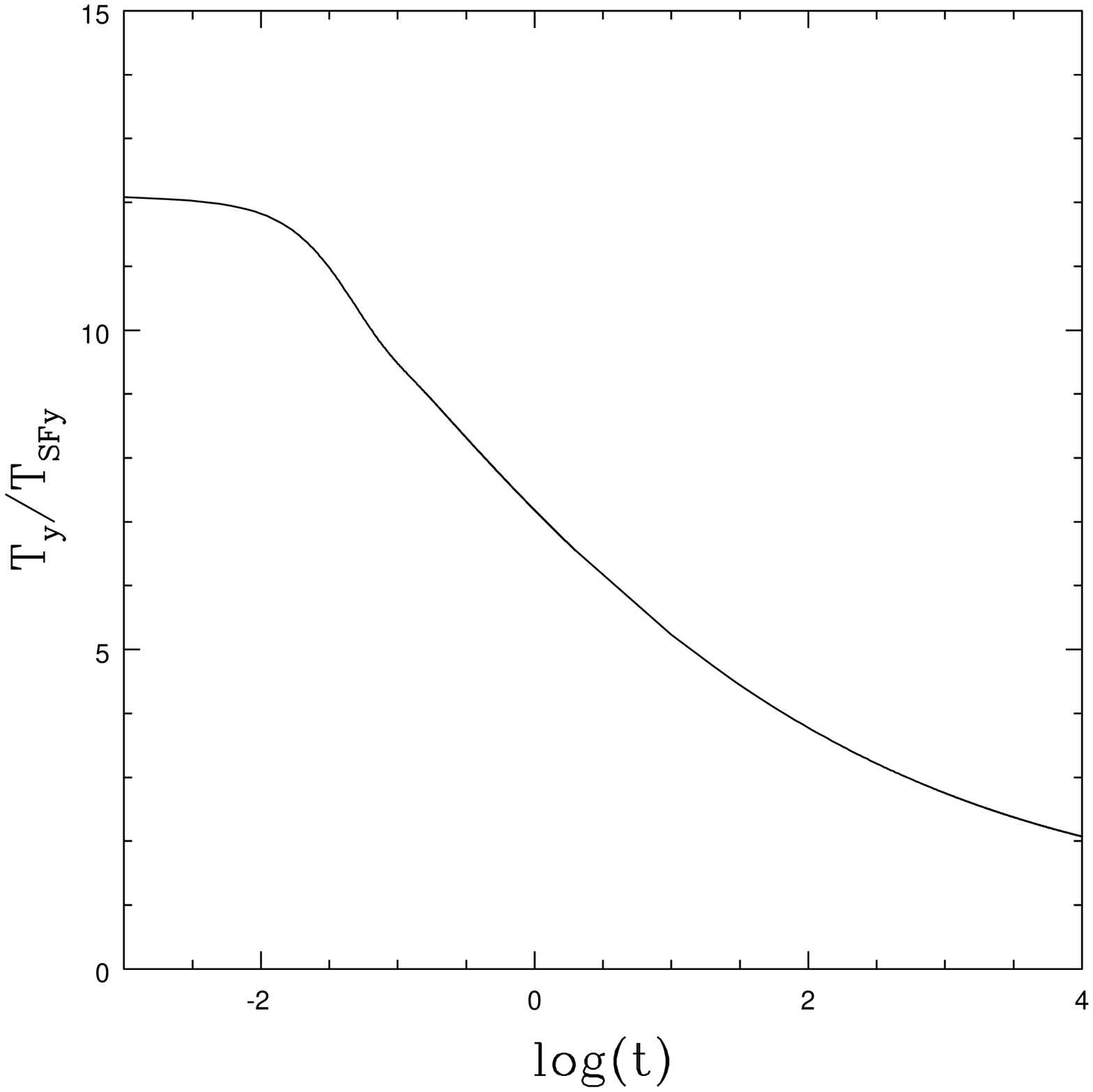,width=0.5\textwidth}}
\caption{Evolution of the the $x$- (left panel) and $y$-component
(right panel) of the torque exerted by the disc on the hole, normalized
to their asymptotic values in a steady state, in the case where
$\nu_2/\nu_1=10$. The $x$-component, responsible for the alignment of
the hole, is never very far from the asymptotic value, except in the
very early stages of the evolution. On the other hand, the
$y$-component, responsible for the precession of the hole's spin can be
significantly different and approaches the steady state value more
slowly.}
\label{fig:torque}
\end{figure*}

\subsection{Validity of our approach and theoretical uncertainties}

Our basic evolutionary equation [Equation (\ref{eq:pringle})] assumes
that the evolution of the disc warp is diffusive and the response of
the disc to the warping perturbation is completely determined, in our
approach, by the coefficient $\nu_2$. The basic motivation for our
assumption stems in the analysis of \citet{pappringle83}, who have
shown that the warp propagation is indeed diffusive (as long as
$\alpha>H/R$) in the linear regime, and assuming that the internal disc
motions are purely hydrodynamic, and that the viscosity present is the
standard Navier-Stokes viscosity. As mentioned above,
\citet{pappringle83} also find that $\nu_2/\nu_1=1/2\alpha_1^2$, which
is generally much larger than unity.

In fact, even with these simplifying assumptions, the linear,
hydrodynamic analysis predicts the presence of precessional terms
\citep{pappringle83,ogilvie00}. These are however second order terms
and we neglect them here.

However, the behaviour of real discs is surely more complex than what
is described simply by Equation (\ref{eq:pringle}).  Turbulence is
likely to be controlled by magnetic fields \citep{balbusreview}, and
what kind of viscosity arises under these conditions is yet to be fully
determined. The only attempt at measuring $\nu_2$ in the presence of
MHD turbulence has been done by \citet{torkelsson00}, who find, within
the limitations of a shearing box approximation, that also in this case
one should expect $\nu_2/\nu_1$ to be relatively large.

Additionally, non-linear effects are going to be important (indeed, as
described below, most of our simulations deal with non-linear
warps). The behaviour of the disc in such regimes is much harder to
describe. The only analytical theories of non-linear warps are by
\citet{ogilvie99,ogilvie00} and \citet{ogilviedubus}. These are for
purely hydrodynamic discs and they generally assume a polytropic
equation of state. In addition, such analyses are limited to mildly
non-linear warps and the resulting non-linear corrections are dependent
on the detailed disc structure ($\Sigma(R)$, $H(R)$, ...). In
particular, \citet{ogilvie99} has shown that the first non-linear
corrections to the diffusion coefficients become important when the
amplitude of the warp $\psi=R|\partial {\bf l}/\partial
R|\gtrsim\alpha$. For these amplitudes, the horizontal shear motions
induced by the warp become supersonic, inducing shocks and additional
dissipation in the disc, not included in \citet{ogilvie99}. In the
presence of very large warps there is also the possibility
\citep{gammie00} of additional dissipation caused by fluid
instabilities. In this case, one possibility is that the ratio
$\nu_2/\nu_1$ might be substantially reduced.

In view of all the above uncertainties, we have opted for a
straightforward, and somewhat abstract, approach. The reason for doing
so is that it keeps the physical processes involved simple, and does
not specialize to any particular disc application (such as AGN, X-ray
binaries, etc.). This enables a basic understanding of the physics
involved in the alignment (or possibly counter-alignment, see below)
process, without adding physically uncertain, and likely unnecessary,
complications.

We have therefore adopted the simple approach of Equation
(\ref{eq:pringle}), but allowing the main parameter $\nu_2/\nu_1$ to
lie within a large range of values (from very large values, as expected
from the linear analysis, to smaller values, perhaps $\approx 1$, that
might arise in strongly non-linear conditions), determining in this way
the dependence of the alignment process on this basic unknown
parameter.

More complex numerical simulations will be needed to determine the
local response of a thin disc in strongly non-linear conditions, but we
leave this for future analyses. We expect that the behaviour determined
here will qualitatively describe the disc evolution (using the
appropriate values of the diffusion coefficient) even in strongly
non-linear regimes.

\section{Steady-state solutions}

\citetalias{scheuer96} have found analytical steady-state solutions to
equation (\ref{eq:pringle}), in the presence of the forced precession
for small tilt angles described in equation (\ref{eq:lense}), subject
to various simplifying assumptions about the disc density distribution
and the inner boundary. The steady-state warp only depends on a single
parameter $R_{\rm w}=\omega_{\rm p}/\nu_2$, the warp radius, which is
the radius at which the warp propagation time-scale equals the local
forced precession rate. Roughly speaking, the disc becomes aligned with
the hole spin at $R\ll R_{\rm w}$ and keeps its original inclination
for $R\gg R_{\rm w}$. The \citetalias{scheuer96} solution takes a very
compact form in terms of the complex variable $W=l_x+il_y$, where $l_x$
and $l_y$ are the $x$ and $y$ component of the unit vector parallel to
the local direction of the disc's specific angular momentum, and the
$z$-axis is parallel to the black hole spin.  In terms of $W$, the
\citetalias{scheuer96} solution reads:

\begin{equation}
W=K\exp\left[-2(1-i)\left(\frac{R_{\rm w}}{R}\right)^{1/2}\right],
\label{eq:sf}
\end{equation}
where $K$ is a complex constant that determines the amplitude of the
warp at large radii. This solution is inaccurate at very small radii,
close to the black hole, and holds only under the assumption that the
inner disc radius $R_{\rm in}\ll R_{\rm w}$ and $K\ll 1$.

Then, \citetalias{scheuer96} compute the torque exerted by the steady
state disc on the black hole and are therefore able to compute the
time-scale over which the hole eventually aligns with the outer disc.
The surface density distribution and the radial profile of the warp are
assumed to correspond to that of a steadily accreting disc and thus
this procedure implicitly assumes that the alignment time-scale is much
longer than the warp propagation time-scale (over which the steady state
solution is attained). The alignment time-scale is roughly given by:
\begin{equation}
t_{\rm align}\sim\frac{J_{\rm h}}{J_{\rm d}(R_{\rm w})} 
\frac{R_{\rm w}^2}{\nu_2},
\label{eq:align1}
\end{equation}
where here $J_{\rm d}(R_{\rm w})$ is the angular momentum of the disc
within the warp radius. The time-scale computed by
\citetalias{scheuer96} is therefore formally valid only if the angular
momentum of the hole is much larger than the angular momentum of the
disc within the warp radius. In the steady-state computed by
\citetalias{scheuer96}, it can be easily shown that:
\begin{equation}
t_{\rm align}=3a\frac{\nu_1}{\nu_2}\frac{M}{\dot{M}}\left(\frac{R_{\rm
s}}{R_{\rm w}}\right)^{1/2},
\label{eq:align2}
\end{equation}
where $a$ is the spin parameter of the hole, $R_{\rm s}$ is its
Schwarzschild radius and $\dot{M}$ is the steady state accretion rate.

A second limitation of the \citetalias{scheuer96} solution, as pointed
out by \citet{king05} (see also Introduction), is that their
steady-state disc formally extends to infinity and therefore has an
infinite total angular momentum. According to \citet{king05}, this
prevents the possibility of obtaining a configuration where eventually
the disc and the hole angular momenta end up being counter-aligned.

With the above discussion in mind, we now assess the validity of the
\citetalias{scheuer96} solution by computing explicitly the evolution
of an initially flat but tilted disc and following its approach to a
steady state.  In this case, we will keep the hole spin direction
fixed. The disc extends from $R_{\rm in}=10^{-2}$ to $R_{\rm
out}=10^2$.  Particular care has been given to the boundary conditions.
In fact, we would like to have a numerical scheme that conserves
angular momentum as far as possible. In order to do this, we implement
the boundary conditions exactly as described in \citet{pringle92}, in
such a way that at both the inner and outer boundary no torque is
provided to the disc in either the azimuthal or the tilt direction
(apart from the Lense-Thirring torques). Note that the usual inner
boundary condition adopted in numerical models of accretion discs
(i.e. $\Sigma=0$, see, for example, \citealt{pringle86}), implies a
zero {\it viscous} torque at the inner edge, thus implying an advective
loss of angular momentum from the inner edge and consequently allowing
accretion. In our implementation, in order to produce an accreting, but
torque-free inner boundary, we set up a sink of material there. In
effect, we add a term on the right-and side of Equation
(\ref{eq:pringle}) of the form:
\begin{equation}
\left.\frac{\partial{\bf L}}{\partial t}\right|_{\rm in}=-\frac{{\bf
L}f(R)}{t_{\rm sink}},
\end{equation}
where $f(R)$ is a dimensionless function strongly peaked at the inner
edge and $t_{\rm sink}$ is an adjustable time-scale, chosen to be small
enough to approximate the usual $\Sigma=0$ inner boundary condition.

 We feed the disc at a constant mass accretion rate $\dot{M}=1$
between $R=70$ and $R=90$, with a flat spatial distribution. Time units
are determined by setting $\nu_1=1$. The initial surface density of the
disc is taken to be that appropriate to a steady state for a flat disc
in this configuration:

\begin{eqnarray}
\label{eq:sigma}
\Sigma_0 &=& (R^{1/2}-R_{\rm in}^{1/2})/3\pi R^{1/2}~~~~~R_{\rm in}<R<R_{\rm
add}\\
\nonumber && (R_{\rm add}^{1/2}-R_{\rm in}^{1/2})/3\pi R^{1/2}~~~~~R_{\rm
add}<R<R_{\rm out},
\end{eqnarray}
where $R_{\rm add}=80$ is the mean radius at which we add matter. The
disc is initially flat, and inclined with respect to the hole spin by
an angle $\theta$. The amplitude of the warp is thus given by
$K=\sin\theta$. Before running the simulations, we have checked that,
in the absence of Lense-Thirring precession, the disc is indeed in a
steady state.

In this case (where the hole spin is kept fixed), the system only
depends on the two parameters $\nu_2/\nu_1$ and $R_{\rm w}$. Below we
discuss separately the two cases where $R_{\rm w}\gg R_{\rm in}$ and
where $R_{\rm w}\approx R_{\rm in}$. Note that we therefore assume that
$R_{\rm in}\gg R_{\rm S}$.

\begin{figure}
\centerline{\epsfig{figure=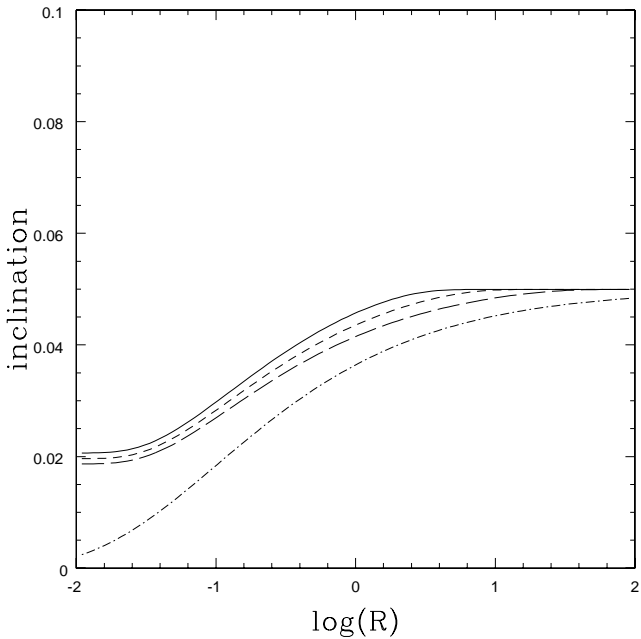,width=0.5\textwidth}}
\caption{Relative inclination between the disc and the hole at $t=0.5$
(solid line), $t=3$ (short-dashed line) and at $t=60$ (long-dashed
line). The steady-state solution in this case is significantly
different than what predicted by \citetalias{scheuer96} (dot-dashed
line). Here $R_{\rm w}=0.025$ ($\log R_{\rm w}=-1.6$).}
\label{fig:steadysmall}
\end{figure}

\subsection{Large warp radius}

In this Section we describe the evolution of the system with a
relatively large warp radius (with respect to the inner disc radius)
$R_{\rm w}= \omega_{\rm p}/\nu_2 = 0.25= 25 R_{\rm in}$. The initial
inclination angle between the disc and the hole was here taken to be
$\theta=0.05$. In Fig. \ref{fig:large} we show the evolution of the
local inclination between the disc and the hole at different times for
the two cases where $\nu_2/\nu_1=1$ (left panel) and $\nu_2/\nu_1=10$
(right panel).  The lines refer to $t=0$ (solid line), $t=0.05$
(short-dashed line) and $t=1$ (long-dashed line), where the times are
expressed in units of the viscous time ($R^2/\nu_1$) at $R=1$. The
dot-dashed line shows the \citetalias{scheuer96} solution in this
case. As expected, in the case where $\nu_2/\nu_1=10$ the propagation
of the warp is much faster so that, at any given time, a larger portion
of the disc is warped in this case with respect to the case where
$\nu_2/\nu_1=1$. Plotting the same quantity as in Fig. \ref{fig:large},
but where the two simulations are portrayed at the same time in units
of $R^2/\nu_2$ (the warp diffusion time-scale), evaluated at $R=1$,
shows that, in this time units, the warped portion of the disc expands
at the same rate in the two cases. 

An interesting feature that can be seen from Fig. \ref{fig:large} is
that in the small $\nu_2$ case the inner disc does not perfectly align
with the hole, as would be predicted by \citetalias{scheuer96}. This is
due to the viscous inward advection of angular momentum from the outer
disc (neglected in the analysis by \citetalias{scheuer96} [cf their
Equation (6) and the following approximations]. This term scales as
$\nu_1/\nu_2$ and indeed the misalignment at small radii disappears in
the right panel in Fig. \ref{fig:large}, where $\nu_1/\nu_2=0.1$.

\begin{figure*}
\centerline{\epsfig{figure=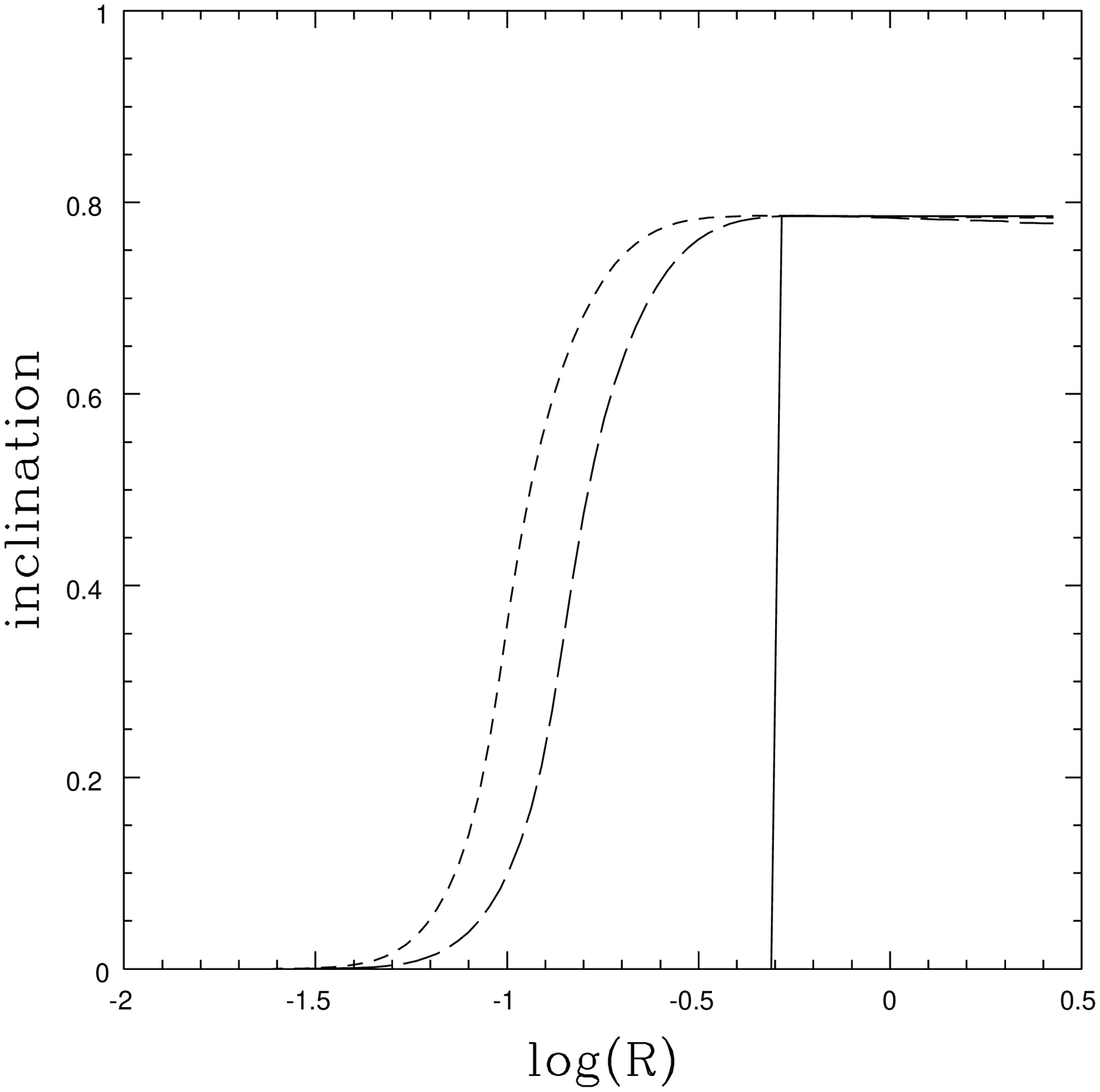,width=0.35\textwidth}
            \epsfig{figure=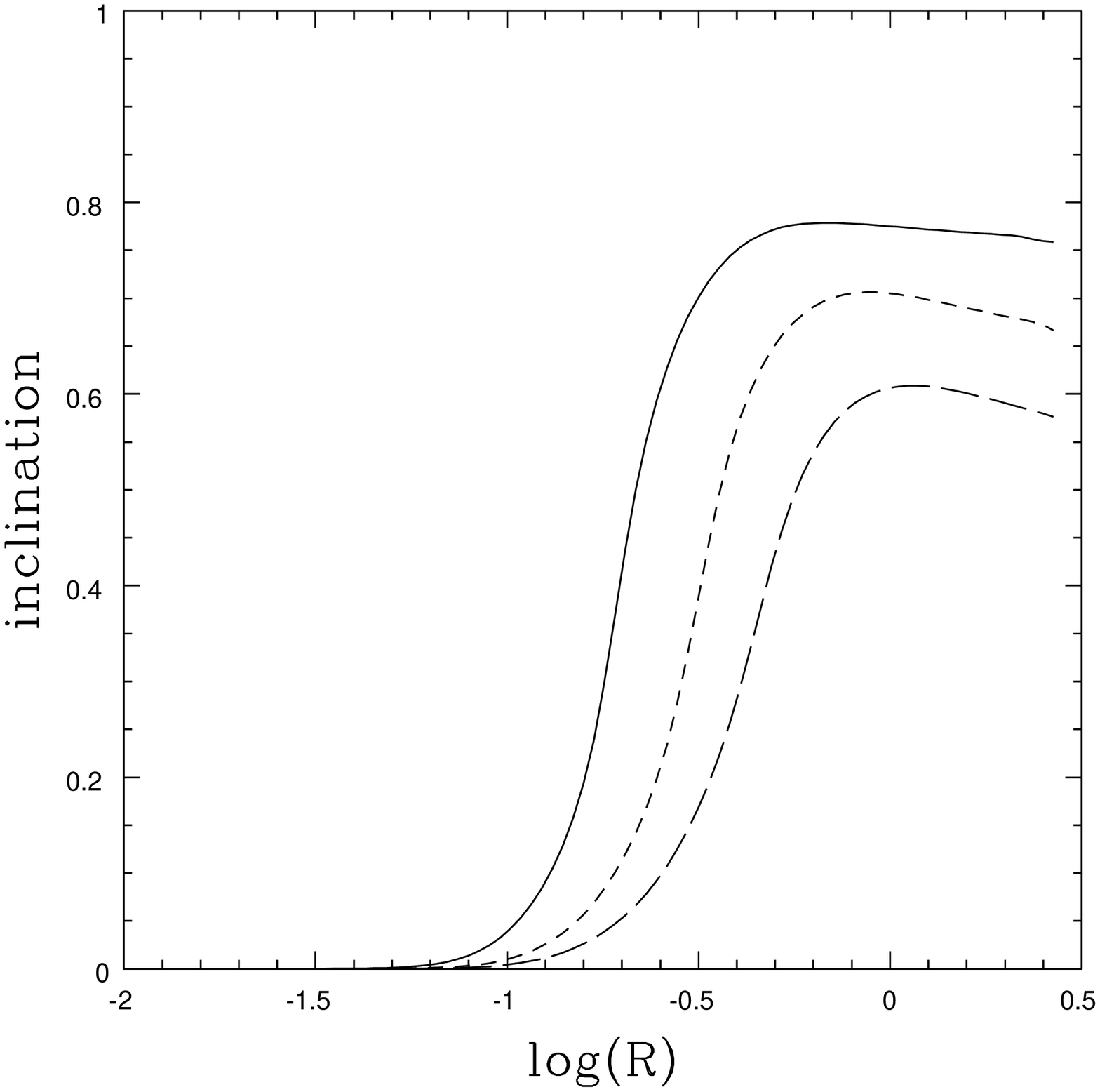,width=0.35\textwidth}
            \epsfig{figure=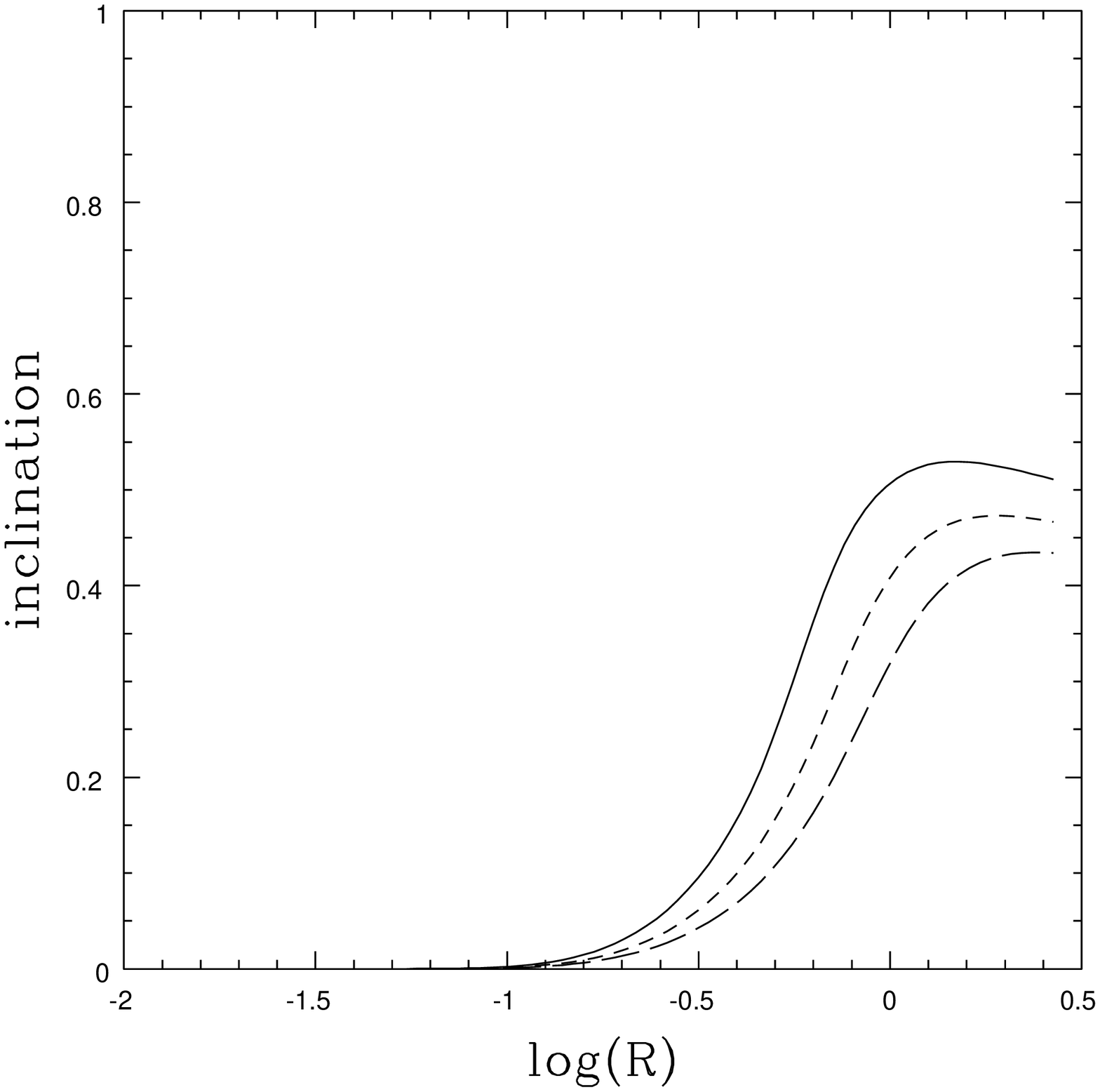,width=0.35\textwidth}}            
\caption{Evolution of the angle between the direction of ${\bf l}(R)$
(the local direction of the disc specific angular momentum) and that of
the black hole angular momentum ${\bf J}_{\rm h}$, at different times
during the evolution, for the case $\log R_{\rm w} = 0.3$ and $J_{\rm
d}=J_{\rm h}$ (Note that throughout the paper we only plot the relative
inclination between the disc and the hole in regions where the disc
surface density is not negligible). Left panel: the lines refer to
$t=0$ (solid line), $t=3\times 10^{-3}$ (short-dashed line) and
$6\times 10^{-3}$ (long-dashed line).  Middle panel: the lines refer to
$t=0.01$ (solid line), $t=0.02$ (short-dashed line) and $t=0.03$
(long-dashed line). Right panel: the lines refer to $t=0.04$ (solid
line), $t=0.05$ (short-dashed line) and $t=0.06$ (long-dashed line). At
$t=0$ the matter is in a ring at $\log R= 0$. By $t\approx 0.3$ the
whole disc has aligned with the hole spin.}
\label{fig:align}
\end{figure*}

An important quantity to be evaluated is the torque ${\bf T}=\de{\bf
J}_{\rm h}/\de t$ exerted on the hole by the disc. Note that, if we
take the $z$-axis to be aligned with the hole's spin, the torque has
only got the $x$ and $y$ components, which are proportional to the $y$
and $x$ components of the disc's angular momentum, respectively. Note
that it is the $x$-component of the torque which is responsible for the
alignment of the hole with the disc's angular momentum, while the
$y$-component gives rise to a precession of the hole spin. In the
\citetalias{scheuer96} steady state, the two components of the torque
${\bf T}_{\rm SF,0}=(T_{{\rm SF}x},T_{{\rm SF}y})$ are equal to each
other in magnitude and are given by:
\begin{equation}
\label{eq:torque}
T_{{\rm SF}x}=-T_{{\rm SF}y}=-\pi K\Sigma_{\infty}\sqrt{\omega_{\rm p}\nu_2GM},
\end{equation}
where $\Sigma_{\infty}$ is the (almost) constant surface density far
from the inner edge of the disc. In our case, we have
$\Sigma_{\infty}=1/3\pi$ (see Eq. \ref{eq:sigma}). In order to check
on which time-scale the asymptotic values of the torques expressed in
equation (\ref{eq:torque}) are attained, we have computed the torques
on the hole exerted by our disc in the case where $\nu_2/\nu_1=10$,
for which our solution eventually approaches the
\citetalias{scheuer96} solution. Note that the values expressed in
equation (\ref{eq:torque}) come from an integral over the disc
surface, assumed to extend from $R=0$ to infinity. For a disc extending
from $R_{\rm in}$ to $R_{\rm out}$ the torque is given by:

\begin{equation}
{\bf T}_{\rm SF}=-2\pi\int_{R_{\rm in}}^{R_{\rm out}}
\frac{\mbox{\boldmath$\omega$}_{\rm p}\times{\bf L}}{R^3} R\de
R=C(R_{\rm in},R_{\rm out}) {\bf T}_{\rm SF,0},
\end{equation}
where $C(R_{\rm in},R_{\rm out})$ is a correction factor of order unity,
that can be easily computed analytically based on the SF solution
[Eq. (\ref{eq:sf})]. 

Fig. \ref{fig:torque} shows the evolution of the $x$- (left panel) and
$y$- (right panel) components of the torque, normalized to the
asymptotic value predicted on the basis of the \citetalias{scheuer96}
solution, $T_{\rm SF}$. Initially, the disc is only inclined with
respect to the hole and does not precess ($l_y$ is initially zero), and
therefore the $x$-component of the torque is also vanishing. In
contrast, the $y$-component of the torque, dependent on the relative
inclination and responsible for the precession of the hole, is much
larger than the asymptotic value. As the evolution proceeds, the disc
acquires a non-vanishing $l_y$ and starts to exert an aligning torque,
$T_x$, on the hole. This reaches a maximum of $\approx 1.5T_{\rm SF}$
at $t\approx 0.05$, and then slowly decreases, approaching the
asymptotic value. Note that the value of $T_x$ is never very far from
the estimate of \citetalias{scheuer96}, which implies that the
time-scale given by SF for the alignment of the hole is roughly
correct. On the other hand, the component $T_y$ of the torque is
significantly different than the \citetalias{scheuer96} solution and
approaches it much more slowly. This means that we expect the hole spin
to precess at a significantly different and apparently faster rate.

\subsection{Small warp radius}

Here we describe the evolution of the system with a smaller warp
radius $R_{\rm w}=2.5R_{\rm in}$. The ratio of the two viscosities is
here taken to be $\nu_2/\nu_1=10$. The initial inclination angle
between the disc and the hole was here taken to be $\theta=0.05$. Fig.
\ref{fig:steadysmall} describes the evolution of the relative
inclination between the hole and the disc with time. The solid line
refers to $t=0.5$, the short-dashed line to $t=3$, while the
long-dashed line refers to $t=60$ (in units of the viscous time-scale
at $R=1$). The disc in this case eventually approaches a steady state,
but this is significantly different than what predicted by
\citetalias{scheuer96} (shown in Fig. \ref{fig:steadysmall} with a
dot-dashed line). This is to be expected, since their solution is only
valid when the warp radius is much greater than the inner disc
radius. Because the warp radius is not much larger than the inner disc
radius, the inner disc is flat, but remains misaligned with the angular
momentum of the hole. 

\section{Warped spreading rings}

From the discussion of previous section, it appears that an initially
flat disc approaches the steady-state warped configuration rather
slowly, on a time scale $t_{\nu_2}=R^2/\nu_2$. The surface density of
the disc, on the other hand, evolves on the viscous time scale
$t_{\nu_1}=R^2/\nu_1$. The ratio of the two time scales is $t_{\nu_2}/
t_{\nu_1}=\nu_1/\nu_2$. Based on the linear calculations by
\citet{pappringle83}, this ratio is small and the warped configuration
can be achieved before the disc is accreted. However, especially for
large warps, the linear calculations are expected to break down and
the two time scales might even become comparable. It might then be
questionable whether the \citetalias{scheuer96} solution can be
actually attained in realistic systems.

In reality, as we have mentioned, the surface density profile of the
disc is unlikely to be that of a steady disc. In practice what we wish
to model is an accretion event (or a succession of events) onto the
central black hole in a galactic nucleus. Such events appear to occur
at random orientations relative to the black hole spin. For this
reason, in order to examine more carefully the dynamical evolution of a
warped disc under relatively more realistic configurations, we have
performed a number of other simulations, where we evolve a system
initially comprising a thin ring of matter orbiting around a black hole
whose spin is misaligned with the direction of the ring rotation.
These simulations have also the advantage of having a well defined
total disc angular momentum. In this case, we also allow the hole spin
to evolve according to equation (\ref{eq:hole}). In this case, we can
use a slightly smaller dynamical range, since throughout the
simulations the ring never spreads outside $R=5$. Therefore, the grid
in this case extends from $R=10^{-2}$ to $R=5$ for most simulations. In
a couple of cases however (when we consider small warp radii, see below
section 4.2) the inner radius was decreased to $R_{\rm in}=5\times
10^{-3}$, in order for it to be significantly smaller than the warp
radius. We initialize the simulations by setting up a thin ring of
matter, with a Gaussian surface density profile, centered around
$R_0=1$ and with an initial width equal to $\Delta R=10^{-2}$. The disc
is initially flat and inclined by an angle $\theta$ with respect to the
hole spin. The modulus of the angular momentum of the disc is therefore
$J_{\rm d}=M_{\rm d} \sqrt{GMR_0}$ (where $M_{\rm d}$ is the total disc
mass). The governing equations only depend on the parameters
$\nu_2/\nu_1$, $J_{\rm h}/J_{\rm d}$, and on the ratio $R_{\rm
w}/R_0$. We have performed a number of different simulations by varying
the parameters above. In the following we will describe in more detail
the behaviour of the system for a number of different initial
configurations.

\begin{figure}
\centerline{\epsfig{figure=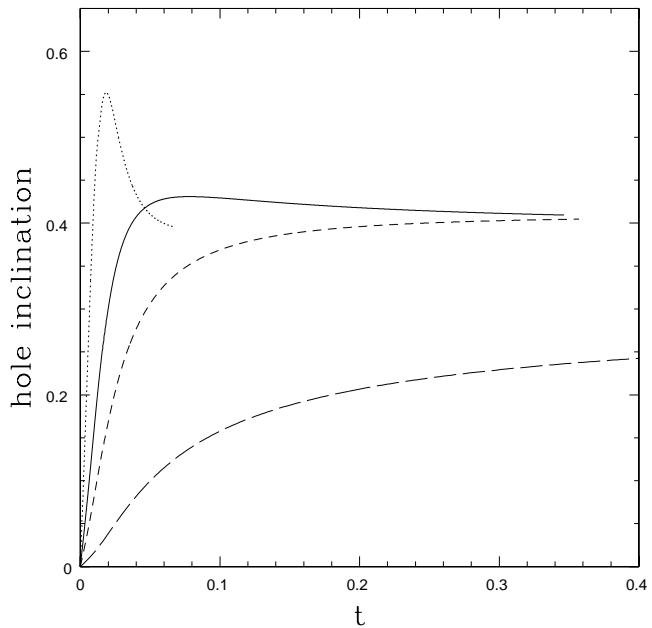,width=0.5\textwidth}}
\caption{Evolution of the angle between the hole spin and the $z$-axis
for the case $J_{\rm d}=J_{\rm h}$ and $R_{\rm w}/R_0=2$. If angular
momentum is conserved, and if the systems becomes eventually completely
aligned, we would expect the final inclination to be $\phi\approx
0.4$. The lines refer to $\nu_2/\nu_1=30,10,5$ and 1 (dotted, solid,
short-dashed and long-dashed lines, respectively).}
\label{fig:hole_align}
\end{figure}

\begin{figure}
\centerline{\epsfig{figure=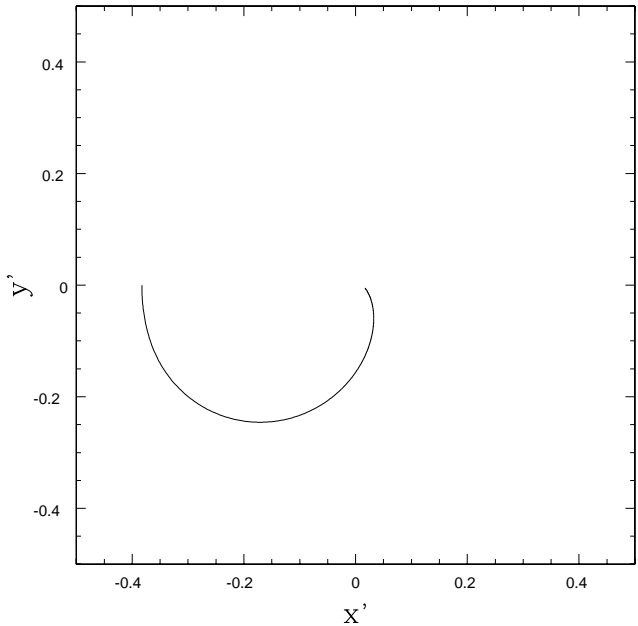,width=0.5\textwidth}}
\caption{The case $J_{\rm d}=J_{\rm h}$, $R_{\rm w}/R_0=2$ and
$\nu_2=10\nu_1$. Evolution of the tip of the unit vector parallel to
the black hole spin instantaneous direction in the $(x',y')$ plane,
perpendicular to the direction of the total angular momentum of the
system ${\bf J}_{\rm h}+{\bf J}_{\rm d}$. The hole aligns after roughly
half a turn in this plane, showing that the precession time-scale is of
the same order of the alignment time-scale.}
\label{fig:prec_align}
\end{figure}

\subsection{Alignment}
\label{sec:large_align}
In this case, we take $R_{\rm w}=2R_0=200 R_{\rm in}$ (so that the
initial material is deposited just outside the warp radius) and
$\nu_2/\nu_1=10$. We also take the initial relative inclination to be
$\theta=\pi/4$ and the ratio of angular momenta to be $J_{\rm d}/J_{\rm
h}=1$. In this case, the condition expressed in inequality
(\ref{eq:king}) would tell us that we expect the disc and the hole to
align.

The evolution of the relative inclination of the disc and the hole is
shown in Fig. \ref{fig:align}. The plots display the angle between the
local direction of the disc specific angular momentum ${\bf l}$ and
the direction of the hole angular momentum, as a function of radius
$R$, at different times during the evolution (see caption for
details), expressed in units of the viscous time-scale at $R=R_0$,
$t_0=R_0^2/\nu_1$. At $t=0$ the disc is flat and confined in a narrow
ring, but then, as it spreads to smaller radii, it also gets warped
and the inner disc tends to become aligned with the hole. By the time
$t\approx 0.3t_0$, most of the disc has aligned with the hole. This is
consistent with the propagation of the warp being roughly a factor
$\nu_2/\nu_1=10$ faster than the spreading of the ring.

The solid line in Fig. \ref{fig:hole_align} shows the evolution of the
angle between the hole spin and the $z$-axis (which coincides with the
initial direction of the hole spin). During the evolution of the
system, angular momentum is not perfectly conserved, because mass (and
correspondingly angular momentum) is being removed from the inner disc,
to allow for accretion. However, the angular momentum removed in this
way is, in general, a small fraction of the total angular momentum,
since the specific angular momentum in the inner disc is a factor
$\sqrt{R_{\rm in}/R_0}\sim 0.1$ smaller than the total. Neglecting this
contribution and assuming that the total angular momentum of the system
${\bf J}_{\rm h} + {\bf J}_{\rm d}$ is conserved, we can get a simple
relationship between the final direction of the hole spin and the
initial relative misalignment. In fact, if the system becomes
eventually completely aligned, all components of the angular momentum
should align with the direction of the total angular momentum (which is
conserved). We then get that the angle $\phi$ between the initial and
final direction of the hole spin is given by:

\begin{equation}
\cos\phi=\frac{1+\xi\cos\theta}{\sqrt{1+\xi^2+2\xi\cos\theta}},
\end{equation}
where $\xi=J_{\rm d}/J_{\rm h}$. In the present case, where
$\xi=1$ and $\theta=\pi/4$, we expect $\phi\approx 0.4$, in agreement
with the result shown in Fig. \ref{fig:hole_align}. 

\begin{figure}
\centerline{\epsfig{figure=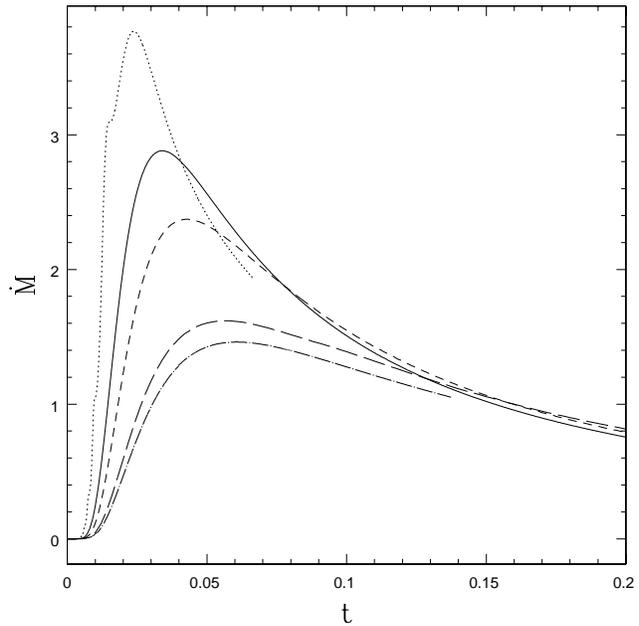,width=0.5\textwidth}}
\caption{The case $J_{\rm d}=J_{\rm h}$, $R_{\rm w}/R_0=2$. Evolution
of the mass accretion rate on the hole in the cases where
$\nu_2/\nu_1=30, 10, 5$ and 1 (dotted, solid, short-dashed and
long-dashed line, respectively), and in the case of a flat spreading
ring (dot-dashed line). The twisting of the disc and the mixing of
angular momenta between different radii increases the accretion rate by
as much as a factor of 3-4.}
\label{fig:mdot_align}
\end{figure}

\begin{figure}
\centerline{\epsfig{figure=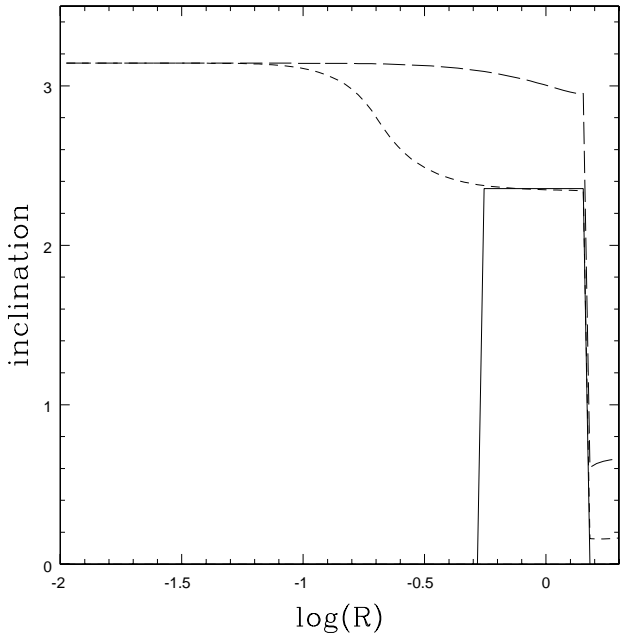,width=0.5\textwidth}}
\caption{Evolution of the angle between ${\bf l}(R)$ and ${\bf J}_{\rm
h}$ in the case where initially $\theta=3\pi/4$ and $J_{\rm d}/J_{\rm
h}=1$, $\log R_{\rm w}=0.3$ and $\nu_2/\nu_1=10$. The system eventually
becomes counter-aligned The lines refer to times $t=$0, 0.01 and 0.1 in
units of $R_0^2/\nu_1$.}
\label{fig:anti}
\end{figure}

During the alignment process, the effect of the torques between the
hole and the disc is not only to align the relative angular momenta,
but, as discussed above, also to produce a precession of both the
angular momentum of different annuli in the disc and of the spin of
the black hole. In order to illustrate this, we plot in Fig.
\ref{fig:prec_align} the evolution of the tip of the unit vector
parallel to the instantaneous direction of the black hole spin in a
plane $(x',y')$ perpendicular to the (constant) direction of the total
angular momentum of the system. At the end of the alignment process,
the hole spin should be parallel to the total angular momentum and the
tip should therefore tend to become very close to the origin of this
plane. This is indeed the behaviour shown in Fig.
\ref{fig:prec_align}. The hole spin starts with a relatively large $x$
component (it starts on the left in this plot) and, after roughly half
a turn in this plane, eventually aligns with the total angular
momentum. This also shows that the precession time-scale is of the
same order as the alignment time-scale in agreement with the
conclusion of \citetalias{scheuer96}.

The solid line in Fig. \ref{fig:mdot_align} shows the evolution of the
mass accretion rate with time.  As we can see, the presence of the warp
results in an enhancement of the mass accretion rate by as much as a
factor 3, with respect to a non-warped spreading ring calculation
(analogous to the standard case discussed by \citealt{pringle81}), here
shown with a dot-dashed line. This is due to the fact that the term
associated with $\nu_2$ induces some energy dissipation in the disc,
which then is driven more efficiently to small radii. This process was
already noted and described by \citet{pringle92}.

In order to illustrate how the behaviour of the disc changes when
different values of $\nu_2/\nu_1$ are used, we have also run a number
of simulations, identical to the previous one, but with
$\nu_2/\nu_1=30$, $\nu_2/\nu_1=5$ and $\nu_2/\nu_1=1$. In actual discs,
the ratio $\nu_2/\nu_1$ might be even larger than our largest value,
but we expect that the trends derived here will hold also for even more
extreme viscosity ratios.

\begin{figure*}
\centerline{\epsfig{figure=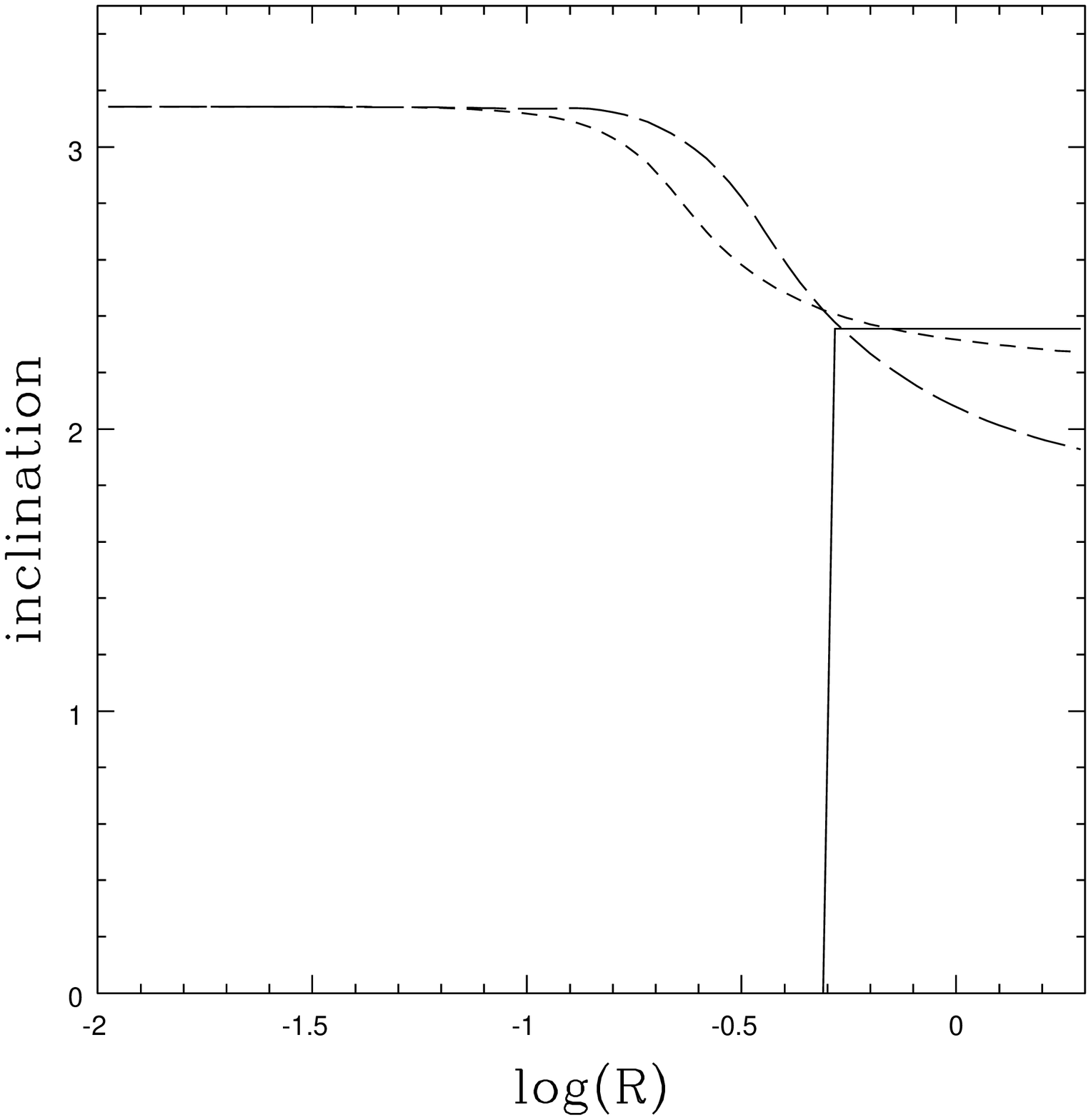,width=0.4\textwidth}
            \epsfig{figure=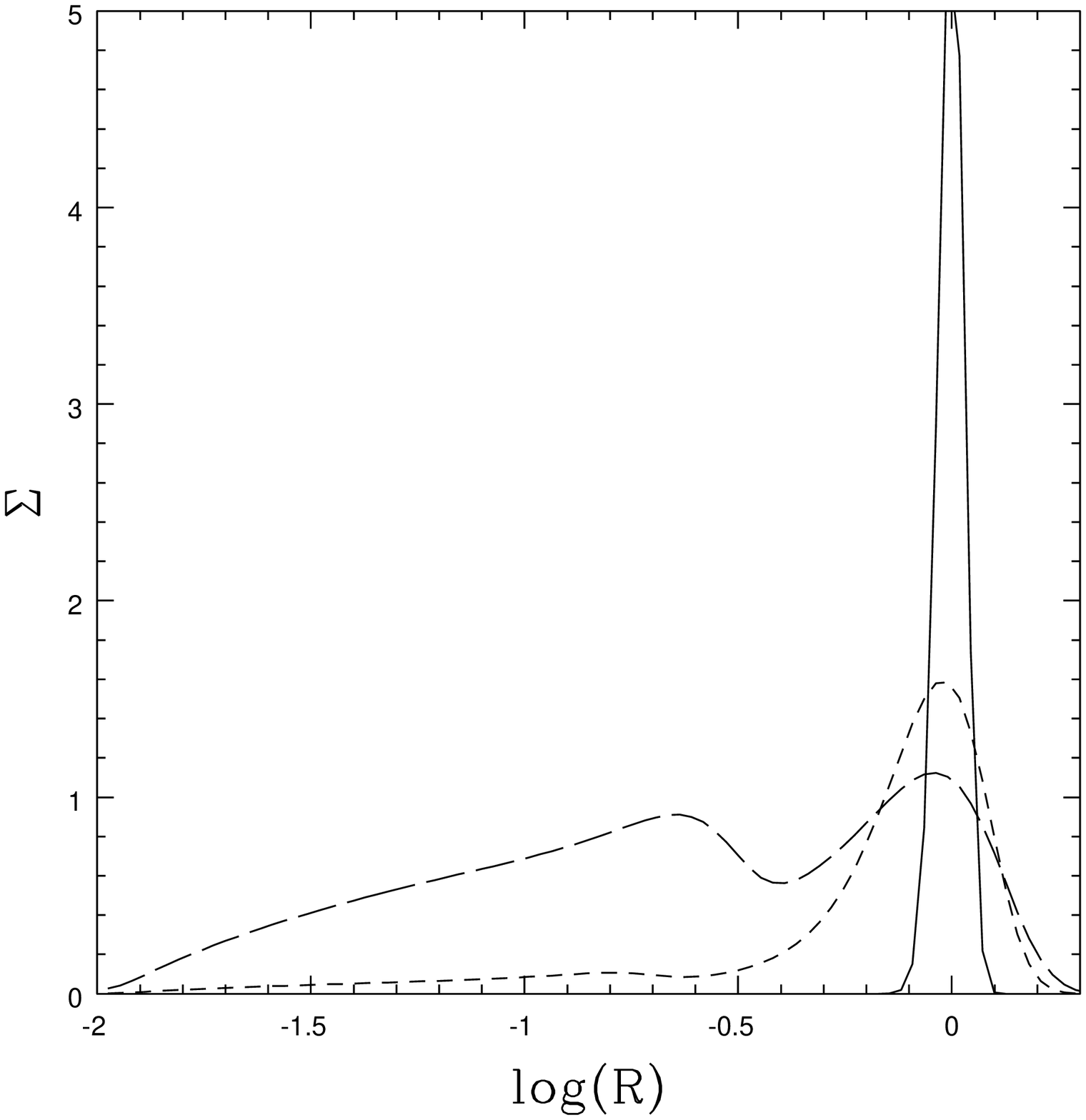,width=0.4\textwidth}}
\centerline{\epsfig{figure=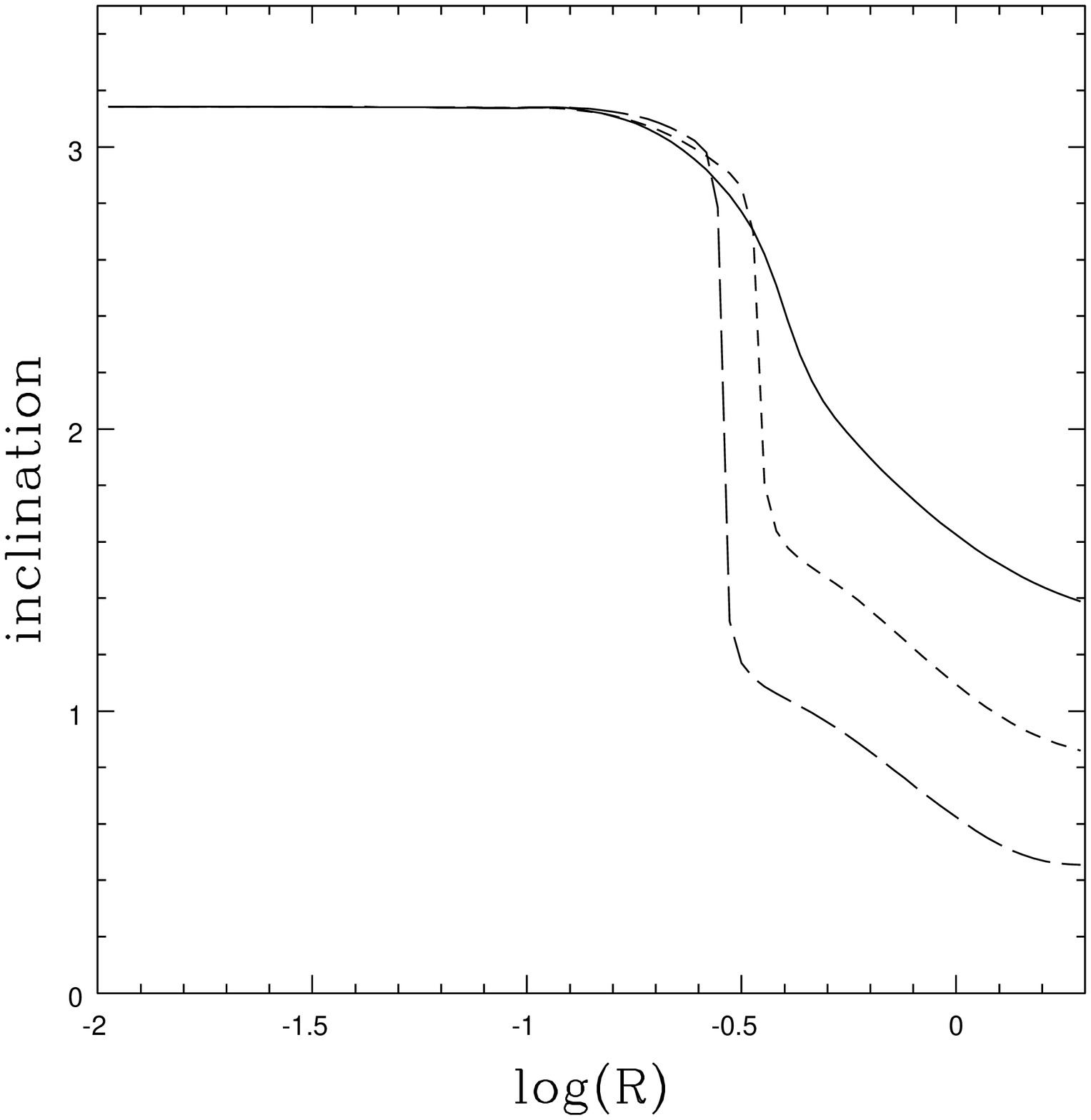,width=0.4\textwidth}
            \epsfig{figure=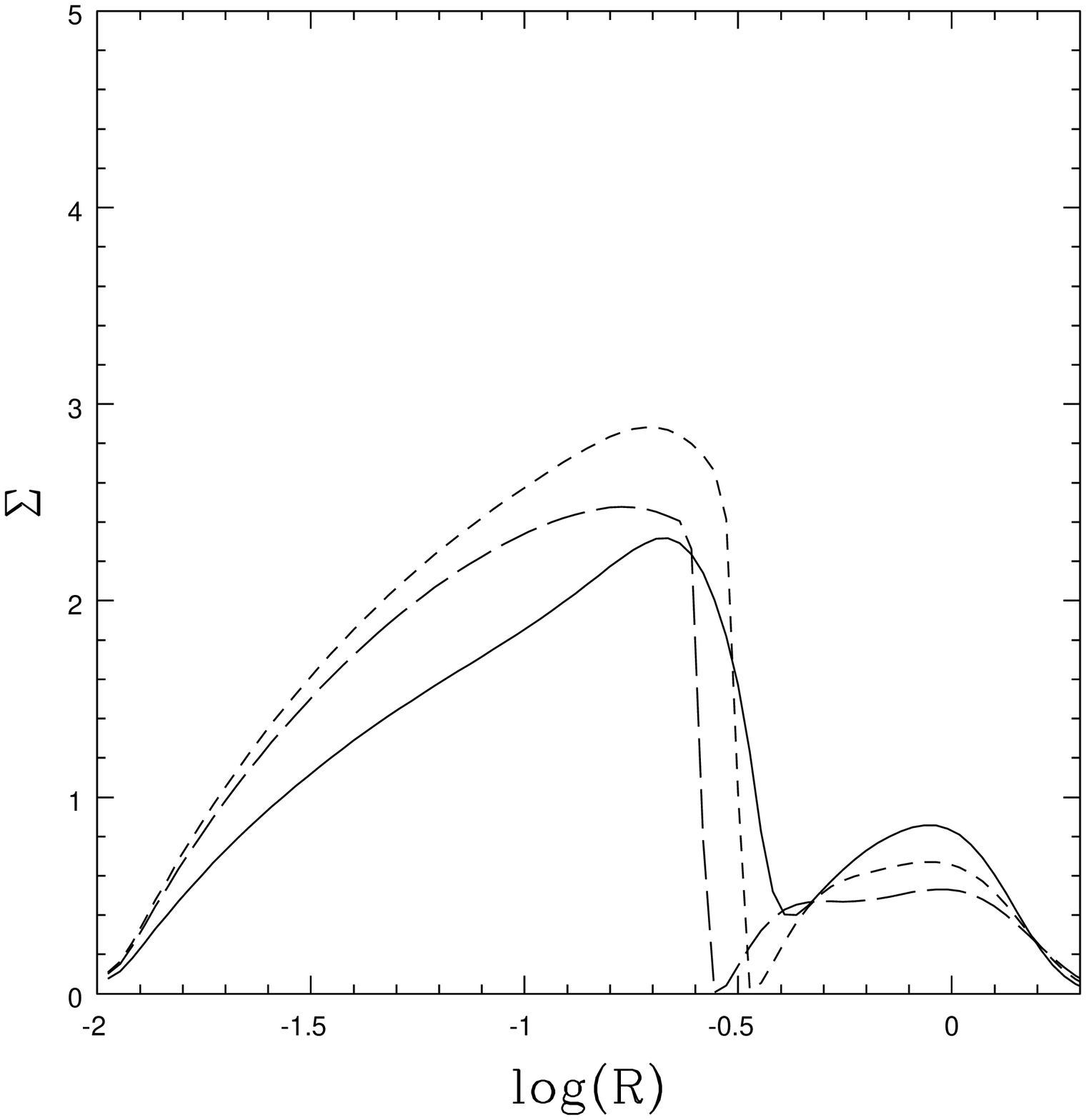,width=0.4\textwidth}} 
\centerline{\epsfig{figure=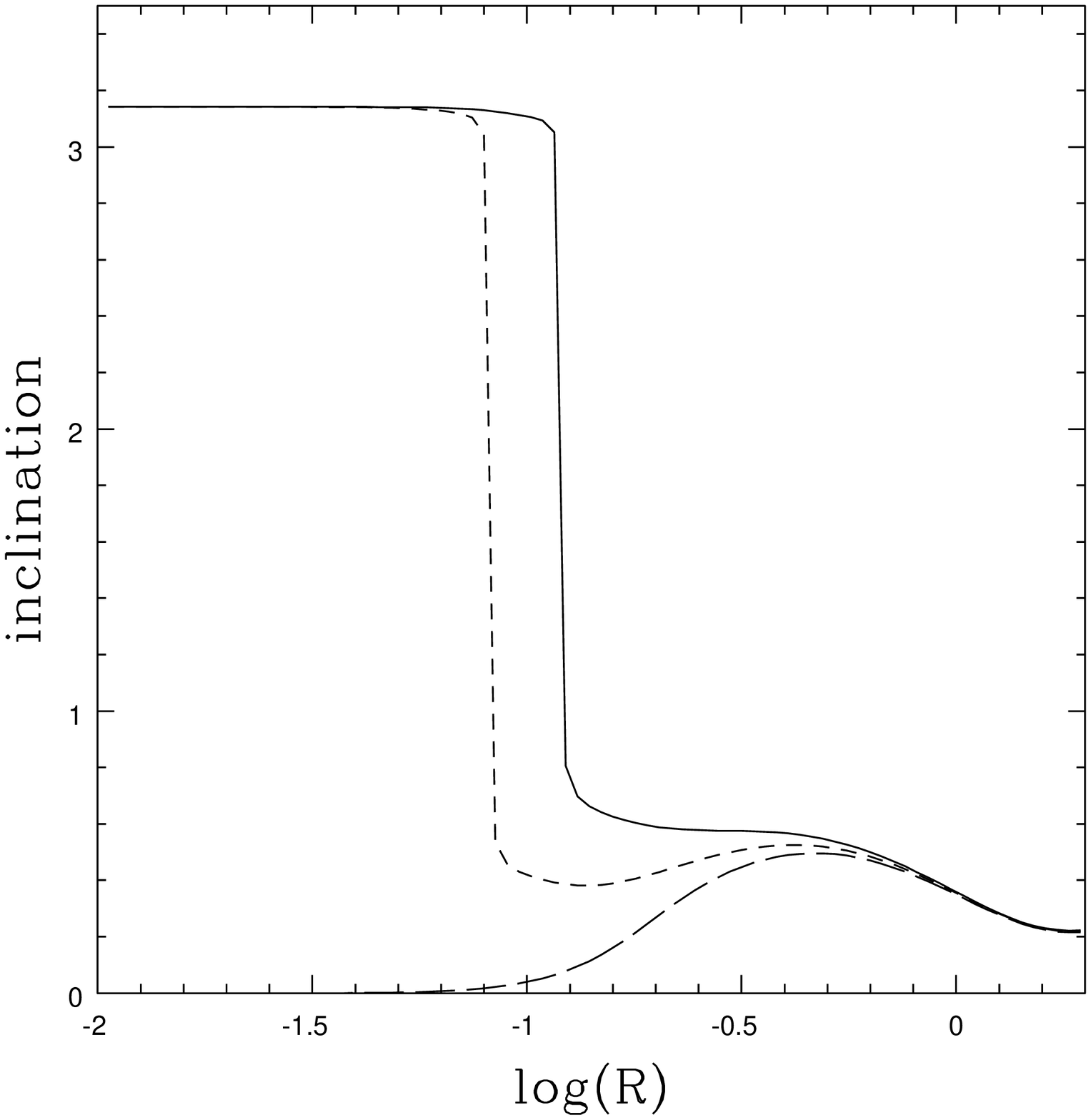,width=0.4\textwidth}
            \epsfig{figure=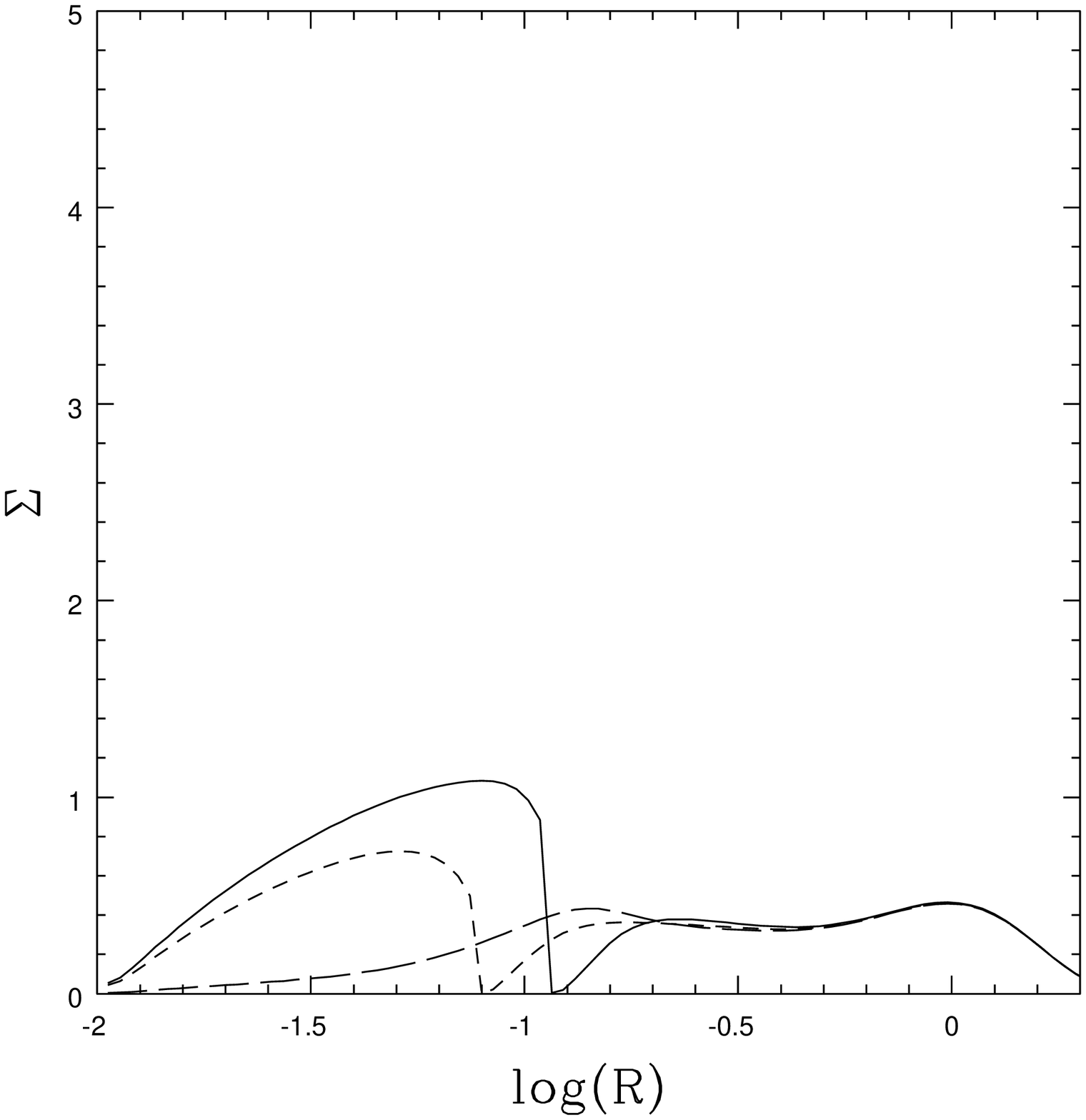,width=0.4\textwidth}}       
\caption{{\bf Left}: Angle between ${\bf l}(R)$ and ${\bf J}_{\rm h}$
at various times during the evolution of a simulation where $J_{\rm
d}/J_{\rm h}=5$ and where the initial misalignment is
$\theta=3\pi/4$. Here $\log R_{\rm w}=0.3$ and $\nu_2/\nu_1=10$. Top
panel : at $t=0$ (solid line), $t=0.01$ (short-dashed line), $t=0.02$
(long-dashed line). Middle panel: at $t=0.03$ (solid line), $t=0.04$
(short-dashed line) and $t=0.05$ (long-dashed line). Bottom panel: at
$t=0.0568$ (solid line), $t=0.0572$ (short-dashed line) and $t=0.0576$
(long-dashed line). Time is in units of $R_0^2/\nu_1$, where the ring
is initially at $R_0=1$. {\bf Right}: Corresponding surface densities
of the disc.}
\label{fig:break}
\end{figure*}

The results of these other simulations are also shown in
Fig. \ref{fig:hole_align} and \ref{fig:mdot_align}, where the dotted
line refers to $\nu_2/\nu_1=30$, the short-dashed line to
$\nu_2/\nu_1=5$ and the long-dashed line to $\nu_2/\nu_1=1$. 

The results shown in Fig. \ref{fig:hole_align} can be easily understood
in terms of Eqs. (\ref{eq:align1}) and (\ref{eq:align2}), according to
which the alignment timescale is inversely proportional to
$\nu_2/\nu_1$. Indeed, when a large $\nu_2/\nu_1$ the disc and the hole
rapidly get aligned between each other. An interesting feature
appearing in the high $\nu_2/\nu_1$ simulations is that the hole
inclination overshoots somewhat before settling down at the asymptotic
value. 

For what concerns the mass accretion rates, as expected, the increase
in accretion when a warp is present is larger for larger values of
$\nu_2/\nu_1$, consistent with our interpretation that it is the
enhanced dissipation associated with $\nu_2$ that drives most of the
accretion process. However, note that the enhanced accretion does not
scale linearly with $\nu_2/\nu_1$. 

\subsection{Counter-alignment}
\label{large_anti}
As a second test, we have considered the case where, as before, $R_{\rm
warp}=2R_0$, $\nu_2/\nu_1=10$ and $J_{\rm d}/J_{\rm h}=1$.  In this
case, however, we start from initially almost anti-parallel angular
momenta, with $\theta=3\pi/4$. In this case \citet{king05} predict that
the disc and hole end-up counter-aligned, since in this case $J_{\rm
d}/J_{\rm h}=1<-2\cos\theta\approx 1.41$. The evolution of the relative
inclination of the disc and the hole is shown in Fig.  \ref{fig:anti}.
The three lines refer to $t=0$ (solid line), $t=0.01$ (short-dashed
line) and $t=0.1$ (long-dashed line). As can be seen the spreading of
the ring is accompanied by a fast counter-alignment of the system (the
angle approaching $\pi$), in agreement with \citet{king05}.

\subsection{Eventual alignment with initial spins almost anti-parallel}
\label{sec:large_rj5}
As we have seen in the previous Section, during the spreading of a ring
of matter towards the black hole, when the initial angular momenta are
almost anti-parallel, the inner part of the disc becomes rapidly
counter-aligned with the spin of the hole. In general, the angular
momentum of the part of the disc that first interacts with the hole is
only a small fraction of the total. Therefore if the disc and hole are
initially close to counter-alignment we would expect that this material
would always counter-align, regardless of the total angular momentum of
the disc. However, as noted by \citet{king05}, when the disc angular
momentum is large, we should expect that eventually the system would
tend to alignment. This brings the question of how this eventual
alignment takes place, when initially the inner disc is expected to be
counter-aligned. In order to answer to this question, we have performed
another simulation, with the following parameters: $R_{\rm w}=2R_0$,
$\nu_2/\nu_1=10$ and $\theta=3\pi/4$. This is similar to the case
discussed above. However, we now take $J_{\rm d}/J_{\rm h}=5$.  In this
case, $J_{\rm d}/J_{\rm h}=5>-2\cos\theta\approx 1.41$, and, according
to \citet{king05}, we should have eventual alignment.

The evolution of the system is shown in Fig. \ref{fig:break}. The left
panels show the evolution of the relative inclination between the disc
and the hole, while the right panels show the corresponding surface
density of the disc. As can be seen the initial spreading is indeed
accompanied by an counter-alignment (top panels). Note the effect of
the enhanced dissipation due to the warp in the evolution of the
surface density. In this case more matter flows in with respect to a
non-warped spreading ring, resulting in a secondary ``bump'' at small
radii in the surface density (compare with Fig. 3 of
\citealt{pringle92}, which shows the same effect, but for a smaller
warp amplitude, and with Fig. 2 of \citealt{pringle81}, which shows the
evolution of a non-warped disc). The middle panels show the evolution
at later stages. The disc has essentially broken into two: an inner
counter-aligned disc and an outer disc, containing most of the disc
angular momentum, with respect to which the system composed of black
hole and inner disc tends progressively to align. The inner disc
becomes smaller in size as time goes by, due to accretion (enhanced by
the warping) and due to the fact that the inner disc is essentially
disconnected from the outer disc.  The bottom plots show the later
evolution of the system. The inner disc is progressively accreted,
while the outer disc becomes progressively more aligned with the hole.
The long-dashed line in the bottom plots shows the final stage of the
evolution, where the inner disc has been completely accreted and we are
left with an aligned disc. The outer parts of the disc, which still
have a small inclination with respect to the hole, are going to be
aligned more slowly, due to the longer time-scales in the outer
disc. We then see that the prediction by \citet{king05} is still valid
in this case.

\begin{figure}
\centerline{\epsfig{figure=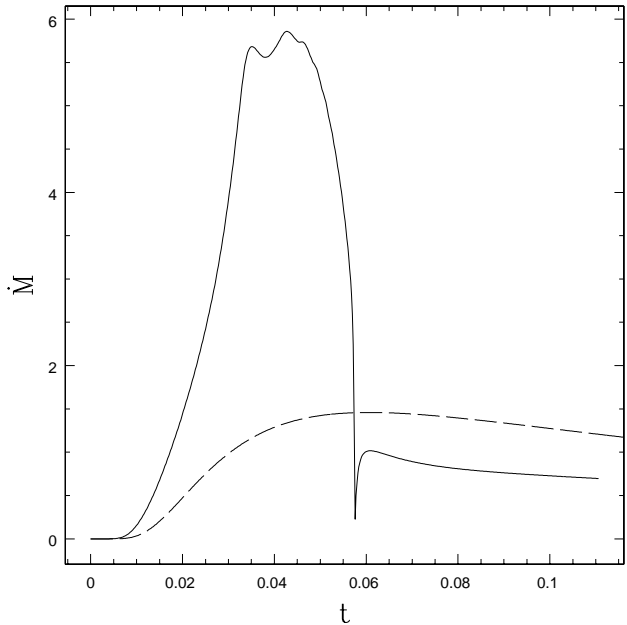,width=0.5\textwidth}}
\caption{Evolution of the mass accretion rate onto the black hole for
the case shown in Figure \ref{fig:break} where initially almost
anti-parallel angular momenta eventually align. The dashed line shows
the corresponding evolution for a flat spreading ring. The accretion
rate is greatly enhanced during the accretion of the inner disc.}
\label{fig:mdotr5}
\end{figure}

\begin{figure}
\centerline{\epsfig{figure=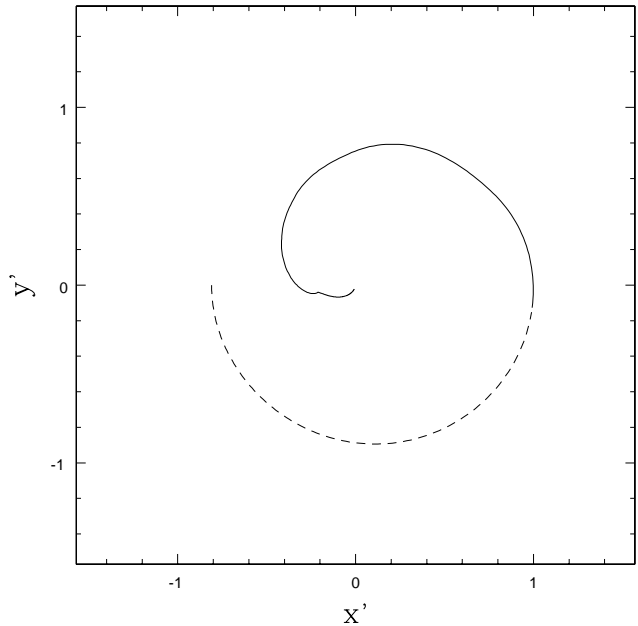,width=0.5\textwidth}}
\caption{Evolution of the tip of the unit vector parallel to the black
hole spin instantaneous direction in the $(x',y')$ plane, perpendicular
to the direction of the total angular momentum of the system ${\bf
J}_{\rm h}+{\bf J}_{\rm d}$, in the case where the initial angular
momentum of the hole is almost anti-parallel with respect to the disc's
one and where $J_{\rm d}/J_{\rm h}=5$ (for details see Figure
\ref{fig:break}). In this case, the hole initially forms an angle
larger than $\pi/2$ with ${\bf J}_{\rm t}$. The evolution is shown with
a dotted line while the angle between ${\bf J}_{\rm h}$ and ${\bf
J}_{\rm t}$ is larger than $\pi/2$, and with a solid line
otherwise. Also in this case, the precession time-scale is of the same
order of the alignment time-scale and the alignment takes place after
roughly one full precession. Note that the black hole spin axis traces
out a large arc on the sky.}
\label{fig:prec_anti}
\end{figure}

The evolution of the mass accretion rate onto the hole is displayed in
Fig. \ref{fig:mdotr5} with a solid line, while the dashed line shows
the corresponding evolution for a flat disc. In this case, the strong
warping produced in the initial counter-aligned configuration leads to a
strong enhancement of the accretion rate, by  factor $\approx 6$. The
accretion rate then suddenly drops when the inner disc has been
completely accreted and then rises again to values of order unity as
the final accretion of the outer disc takes place. During the first
burst of accretion, roughly 20\% of the disc mass is accreted. 

Figure \ref{fig:prec_anti} shows the evolution of the tip of the unit
vector parallel to the hole spin in the $(x',y')$ plane, perpendicular
to the direction of the total angular momentum ${\bf J}_{\rm t}$. In
this case, the initial angle between ${\bf J}_{\rm h}$ and ${\bf
J}_{\rm t}$ is larger than $\pi/2$, even though at later stages the two
vectors eventually align. In Fig. \ref{fig:prec_anti} we display with a
dotted line the early evolution of the spin direction, during which it
has an angle larger than $\pi/2$ with respect to ${\bf J}_{\rm t}$, and
with a solid line the later evolution, when the relative angle is
smaller than $\pi/2$. The hole spin makes a full turn in this plane,
showing that also in this case the precession rate is comparable to the
alignment rate. In the course of the alignment the black hole spin axis
traces out a large arc on the sky.

\subsection{Dependence on warp radius}

One important parameter in the dynamics of the system considered here
is the relative location of the warp radius $R_{\rm w}$ with respect to
the radius within which most of the disc angular momentum resides, here
identified with $R_0$. It is not simple to estimate the relative
location of these two scales for realistic models of AGN discs. If we
take a typical AGN model, such as the one by \citet{collin90} (see also
\citealt{king05}), we find that for a typical choice of parameters the
radius at which the disc angular momentum equals the angular momentum
of the hole is a few times the warp radius. However, one should keep in
mind that these estimates are based on steady-state models of AGN
discs, which might not be applicable on the time-scales considered here
(and are certainly not appropriate in our spreading ring calculations).

In order to check the effect of the warp radius on the alignment
process, we have performed a number of other simulations, similar to
those presented above, but with different warp radius. In this cases,
we have always taken $\nu_2/\nu_1=10$. 

\begin{figure*}
\centerline{\epsfig{figure=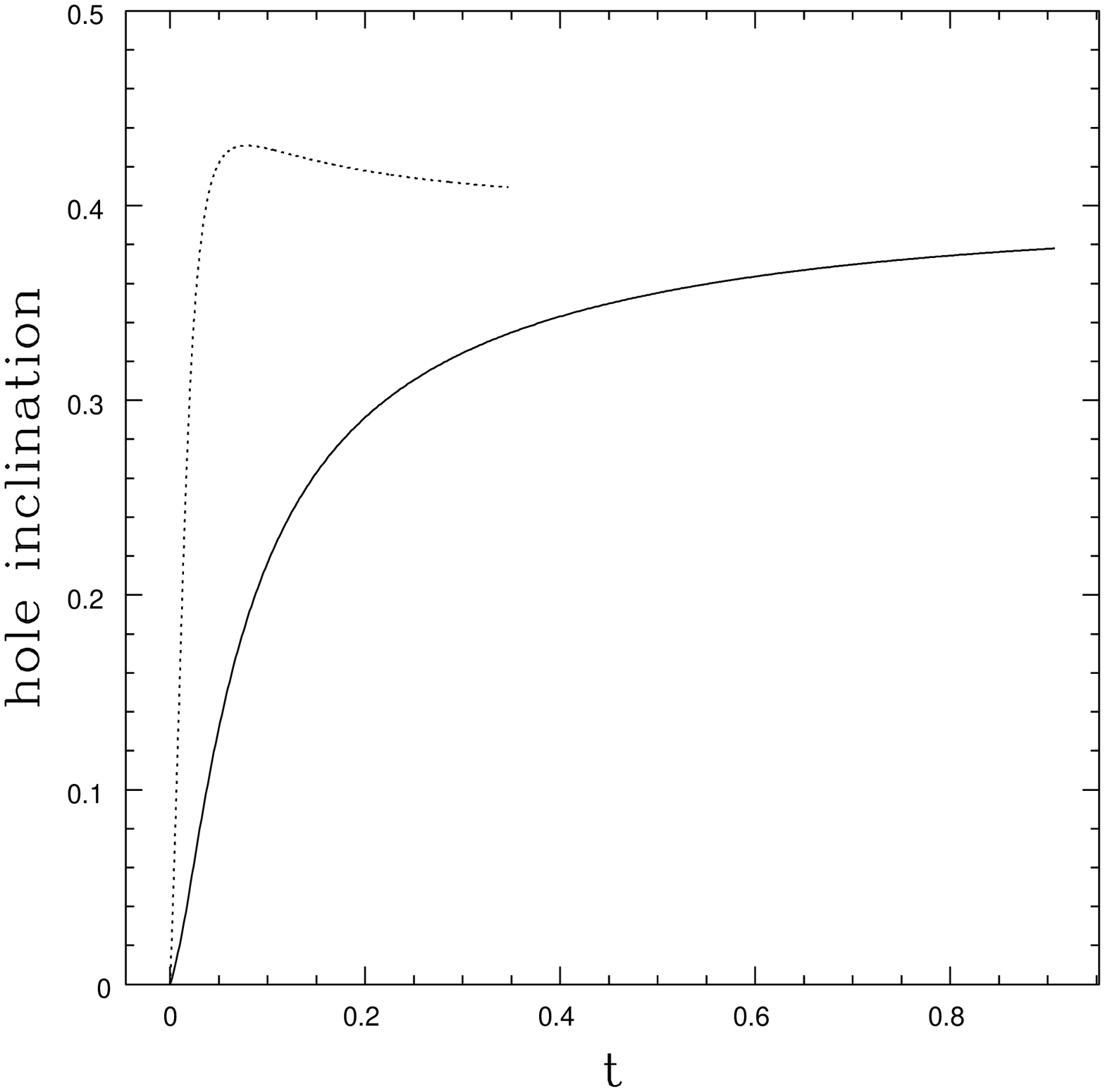,width=0.42\textwidth}
            \epsfig{figure=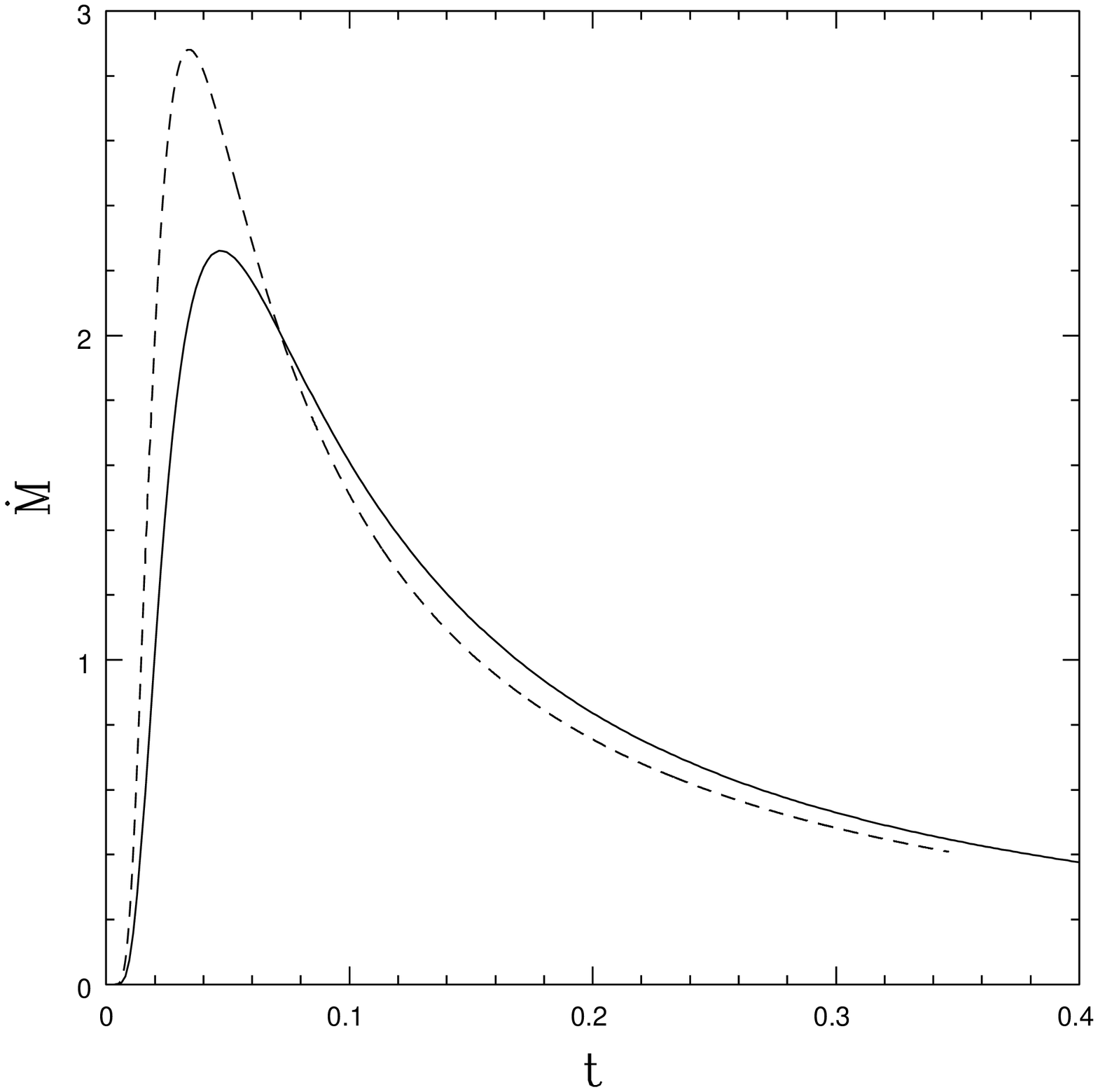,width=0.42\textwidth}}
\caption{{\bf Left}: Evolution of the inclination of the hole spin in
the case where the initial misalignment is $\theta=\pi/4$, $J_{\rm
d}/J_{\rm h}=1$ and $R_{\rm w}=0.25R_0$ (solid line) and $R_{\rm
w}=2R_0$ (dashed line). In the small warp radius case, the evolution of
the hole spin is slower [cf. equation (\ref{eq:align2})]. {\bf Right}:
Mass accretion rate on the hole for the two cases.}
\label{fig:small_align}
\end{figure*}

First, we considered the case where the initial inclination is
$\theta=\pi/4$ and $J_{\rm d}/J_{\rm h}= 1$, in which case we would
expect the disc to align with the black hole (see section
4.1). However, differently than what discussed in section 4.1, we take
here $R_{\rm w}=0.25R_0$, so that the ring is initially completely
outside the warp radius. Figure \ref{fig:small_align} shows the
evolution of hole inclination (left) and mass accretion rate (right)
for this case. The dotted lines show the respective results for the
analogous case with a large warp radius, described above in section
\ref{sec:large_align}. It can be seen that the time-scale for the
alignment of the hole is longer in the case where the warp radius is
smaller, in agreement with equation (\ref{eq:align2}). The peak mass
accretion rate on the hole is smaller in this case, with respect to the
case with a larger warp radius.

Secondly, we considered the case where the initial inclination is
$\theta=3\pi/4$, $J_{\rm d}/J_{\rm h}=1$ and $R_{\rm w}=0.25R_0$. In
this case, see section 4.2, we expect the disc and the hole to
counter-align.  Also this case does not present significant differences
with respect to the case with a larger warp radius. The disc and the
hole counter-align with respect to each other. Again, with a smaller
warp radius the alignment time-scale is longer, as already noted above.

Thirdly, we considered the more interesting case where initially almost
anti-parallel spins end up aligned. This could be a typical situation
in many AGN. We have seen in the previous Section that in this case,
initially the low angular momentum material that first arrives close to
the hole tend to counter-align with the hole but then, after the
accretion of this inner counter-aligned disc, the hole and the outer
disc eventually align. We have performed a number of simulations in
order to check the dependence of the time spent accreting in a
counter-aligned fashion on the choice of the warp radius $R_{\rm
w}$. In all these cases the initial inclination is $\theta=3\pi/4$ and
$\nu_2/\nu_1=10$, but the disc angular momentum is chosen so as to
dominate on the hole, $J_{\rm d}/J_{\rm h}=5$. We have then run a
number of simulations with different $R_{\rm w}/R_0=0.25$, 0.1 and
0.05. In the last two cases, we have decreased the inner disc radius to
$R_{\rm in}=5\times 10^{-3}R_0$, in order for it to be significantly
smaller than the warp radius.  Fig. \ref{fig:incl} shows the
inclination of the inner disc relative to the hole as a function of
time for the simulation where $R_{\rm w}/R_0=$ 0.05 (solid line), 0.1
(dotted line), 0.25 (short-dashed line) and 2 (long-dashed line). It
can be seen that in all cases the systems switches from a
counter-aligned to an aligned configuration, the transition
(corresponding to the accretion of the inner disc, see previous
Section) being progressively more gentle as the warp radius
decreases. This is because, when $R_{\rm w}$ is smaller, the alignment
between the hole and the outer disc is slower [see
Eq. (\ref{eq:align2})], so that when the inner disc is accreted, the
outer disc and the hole are still not aligned. As the warp radius
decreases, the time spent accreting in an counter-aligned fashion
increases and it can be a significant fraction of the accretion
time-scale, up to $t\approx 0.2 t_{\nu_1}(R_0)$, where $t_{\nu_1}(R_0)$
is the viscous time at $R=R_0$. This behaviour can be understood
qualitatively in the following way. We expect that the transition
between the counter-aligned and the aligned configuration occurs
roughly when the angular momentum accreted through $R_{\rm w}$ becomes
comparable with that of the hole [cf. Eq. (\ref{eq:align1})]. Since the
specific angular momentum scales as $R^{1/2}$, in the case where
$R_{\rm w}$ is smaller, a larger amount of matter needs to be accreted
before the accreted angular momentum becomes comparable to that of the
hole, so that the transition consequently occurs later. Equivalently,
the same conclusion can be obtained if one considers that the alignment
time-scale increases with decreasing $R_{\rm w}$
[Eq. (\ref{eq:align2})].

\section{Discussion and conclusions}

In this paper we have considered the evolution of a system comprising
an accretion disc and a central spinning black hole, in the case where
the hole spin is misaligned with the rotation direction of the disc,
and the disc is therefore subject to Lense-Thirring precession and
becomes twisted and warped. Our approach is similar to the one
considered by \citet{pringle92}, in that we consider the case where the
warp propagates in the disc in a diffusive manner. However, while
\citet{pringle92} considered the warping of an otherwise steady-state
disc, here we also consider the case where the surface density of the
disc evolves as the disc is being warped. Furthermore, in our evolution
scheme, the direction of the spin of the hole is allowed to move under
the influence of the torques exerted by the disc.

We have carried out a series of simulations under different conditions.
First, we have considered the warping of a disc which is initially
characterized by a steady-state surface density distribution
\citep{pringle92} and where the direction of the hole spin is kept
fixed and considered various cases, where the location of the warp
radius $R_{\rm w}$ and the ratio of the two viscosities $\nu_2/\nu_1$
are changed. We find that the disc eventually settles down in a steady
warped configuration (similar to the one obtained analytically by
\citetalias{scheuer96}), but this occurs over the relatively long
time-scale associated with the warp propagation time-scale
$t_{\nu_2}\approx R^2/\nu_2$, that can be a significant fraction of the
accretion time-scale. We find that if the warp radius is small and the
ratio of the two viscosities is of order unity, at small radii the warp
structure can be significantly different from that predicted by
\citetalias{scheuer96}.

We also find that the time-scale on which the black hole spin and the
disc align can be comparable with the propagation time for the warp. We
therefore also considered the probably more realistic situation in
which a ring of matter is accreted by the hole (as can happen, for
example, during an accretion event onto a central black hole in a
galactic nucleus). As the ring spreads, it becomes warped and the
direction of the hole spin is modified. A general feature of these
warped spreading ring simulations is that the dissipation associated
with the twisting of the disc leads to an enhanced accretion rate onto
the hole (by as much as a factor 10).

We have computed the detailed evolution of the direction of the spin of
the hole during the accretion and alignment process. We have found, in
agreement with \citetalias{scheuer96}, that the ``precession''
time-scale (i.e. the time-scale of the change of the direction of the
spin perpendicular to the total angular momentum) is comparable to the
alignment time-scale, so that during alignment the spin of the hole
only ``goes round'' at most once (see Figs. 7 and 11). If the jets
emanating from AGN point in the instantaneous direction of the hole
spin, our conclusions have important consequences on the shape of the
observed ``precessing'' jets in these systems. Recently,
\citet{lister03} have reported observations of a ``precessing'' jet
most likely emanating from an AGN. They have fitted the shape of the
observed jet by using a model where the nozzle precesses slowly around
a fixed axis. However, if the perturbation of the axis of the nozzle is
indeed due to changes in the spin orientation of the black hole induced
by the Bardeen-Petterson effect, then our results indicate that the
assumption of precession around a fixed axis might not be correct. On
the other hand, \citet{lister03} estimate that the typical length-scale
of the jet is $\approx 280$ pc, and that the advance speed of the lobes
is $\approx 0.3c$. This would imply a ``precession'' time-scale of
$\approx 3000$ yrs, which is too small to be attributed to the
Bardeen-Petterson effect [cf. our Eq. (\ref{eq:align2})]. In this
particular case, the precession is more likely to result from
spin-orbit coupling in a binary black hole system.

\begin{figure}
\centerline{\epsfig{figure=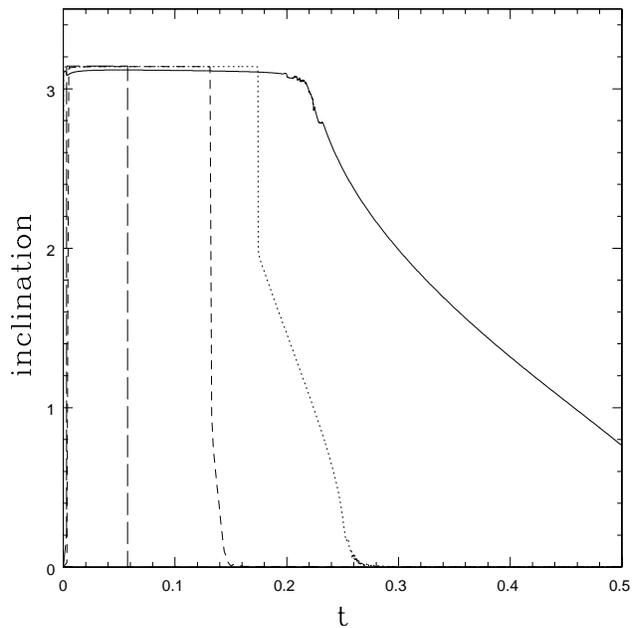,width=0.5\textwidth}}
\caption{Inclination between the inner disc and the hole as a function
of time (in units of the viscous time-scale $R^2/\nu_1$ at $R=R_0$ for
the cases where the initial spins are almost anti-parallel, the angular
momentum ratio is $J_{\rm d}/J_{\rm h}=5$ and $R_{\rm w}/R_0=0.05$
(solid line), 0.1 (dotted line), 0.25 (short-dashed line) and 2
(long-dashed line).}
\label{fig:incl}
\end{figure}

We would like to stress that the conclusion that alignment time-scale
and ``precession'' time-scale are of the same order for the
Bardeen-Petterson effect implies that ({\it i}) if proper precession
around a fixed axis is observed, then it is most likely not due to
Bardeen-Petterson alignment between the disc and the black hole and
({\it ii}) it provides an observational tool to estimate the alignment
time-scale, by looking at the ``precession'' time-scale, for those
systems where we are indeed looking at the Bardeen-Petterson effect. 

We have also considered in detail the issue of the conditions under
which the disc and the hole can be counter-aligned. We have confirmed
the suggestion by \citet{king05} that, if the initial misalignment
between the disc and the hole is larger than 90 degrees and the hole
angular momentum dominates the system, accretion proceeds in a
counter-aligned fashion. However, also in the case where the disc
angular momentum dominates and eventual alignment occurs, a large
fraction of the mass is initially accreted in a counter-aligned
fashion. We have shown (see Fig. \ref{fig:incl}) that the amount of
time spent accreting in a counter-aligned fashion increases as the warp
radius decreases. Indeed, the alignment time-scale
[cf. Eq. (\ref{eq:align2})] scales as $R_{\rm w}^{-1/2}$. In a typical
AGN, the alignment time-scale can be obtained combining our equation
(\ref{eq:align2}) with Eq. (22) of \citet{king05} (to estimate $R_{\rm
w}/R_{\rm S}$). This gives:

\begin{eqnarray}
\nonumber t_{\rm align}\approx 8.1\times 10^6a^{11/16}
\left(\frac{\epsilon}{0.1}
\right)^{7/8} \left(\frac{L}{0.1L_{\rm E}}\right)^{-7/8} \\
  M_8^{-1/16}
\left(\frac{\alpha_1}{0.03}\right)^{15/16} \left(\frac{\alpha_2}{0.3}
\right)^{-11/16} \mbox{yrs},
\end{eqnarray}
where $\epsilon$ is the efficiency of accretion, $L$ is the AGN
luminosity, $L_{\rm E}$ is the Eddington luminosity, $M_8$ is the black
hole mass in units of $10^8M_{\odot}$, and $\alpha_1$ and $\alpha_2$
are the $\alpha$ parameters corresponding to $\nu_1$ and $\nu_2$,
respectively. We therefore see that the alignment time-scale is of the
order of the lifetime of a typical AGN ($\approx 10^7$ yrs), and we
might expect AGN to spend a relatively long time accreting in a
counter-aligned fashion. This counter-aligned accretion mode can have
important consequences on observables of AGN, such as the shape of
X-ray iron lines, since for counter-aligned accretion the last stable
orbit is at a significantly larger radius than for the aligned case.

However, we have also shown that the disc can suffer an abrupt change
in orientation at a radius close to the warp radius, especially during
counter-aligned accretion (see Figs. 4 and 9). A large fraction of the
solid angle as seen from the center will be filled by the disc
itself. This can lead to significant absorption of X-rays produced in
the inner disc from the outer disc \citep{phinney89}. The surface
density of a steady-state AGN disc at $R_{\rm w}$ is \citep{collin90}: 

\begin{eqnarray}
\nonumber \Sigma \approx 2.8\times 10^5 a^{-3/8}
\left(\frac{\epsilon}{0.1}\right)^{-3/4} \left(\frac{L}{0.1L_{\rm
E}}\right)^{3/4} M_8^{1/8}\\
\left(\frac{\alpha_1}{0.03}\right)^{-7/8}
\left(\frac{\alpha_2}{0.3}\right)^{3/8} \mbox{g/cm}^2,
\end{eqnarray}
which is high enough to obscure visibility of the central object. This
implies that direct observation of counter-aligned accretion in
observationally biased against, so that most of those AGN for which
the central regions are visible are most likely to be co-rotating.  

Our results are also relevant in order to determine the black hole spin
history during the earlier phases (which appear to occur at redshift
$1<z<5$) when the black hole acquires most of its mass. During this
phase, each accretion event increases the black hole mass significantly
($\Delta M/M\sim 1 -3$, \citealt{volonteri05}). In general, accretion
of counter-aligned material spins the hole down. However, since the
mass accreted during each event is of the order of the hole's mass,
{\it if all the accreted material rotates about the same axis}, we
would be in a situation like our Fig. 9 and the black hole would end up
aligned and maximally spinning, as suggested by \citet{volonteri05}.
However, if this is not the case, we might regard one accretion event
as a sequence of smaller accretion events where at every time the
angular momentum of the accreted material is small compared to the
hole. In such a situation, most of the accretion can occur in a
counter-aligned way and the hole need not end up maximally rotating.
Thus the result would be a spread of black hole spins (cf Fig. 8 in
\citealt{volonteri05}). We conclude that the outcome for the spin
history during this phase depends crucially on the details of the
merging process.

We should mention that in our approach we have neglected the evolution
of the magnitude of the spin of the hole as a consequence of
accretion. This, however, is a small fraction of the total angular
momentum of the system, since it scales approximately as $M_{\rm
acc}R_{\rm in}^{1/2}$, where $M_{\rm acc}$ is the mass accreted by the
hole and $R_{\rm in}$. We should also mention that in our simulations
we have assumed that $R_{\rm in}=0.005-0.01 R_0$, where $R_0$ is the
initial radius at which matter in injected. In actual AGN discs, the
inner disc radius (which is of the order of the Schwarzschild radius)
might be much smaller than this, but this does not affect our
conclusions as long as $R_{\rm w}$ is much larger than $R_{\rm in}$.

A further important uncertainty in what we have described in this paper
is the magnitude of the viscosity $\nu_2$ associated with the
propagation of the warp. This determines the location of the warp
radius and the time-scale of the alignment
process. \citet{pappringle83} have shown that in the linear regime
$\nu_2/\nu_1=1/2\alpha_1^2$, where $\alpha_1$ is the standard viscosity
parameter. However, it is not yet clear what would happen in the
strongly non-linear case described here. Numerical simulations up to
now have only addressed the case where the disc is relatively thick and
warp propagation occurs in a wave-like manner
\citep{nelson99,nelson00,fragile05}. One possibility is that the flow
becomes unstable \citep{gammie00}, such that perhaps $\nu_2\approx
\nu_1$. In some of the cases discussed here the disc can actually break
in the sense that the tilt angle shows a sharp jump at some radius (see
our Fig. 9, for example). In such conditions (see also the numerical
simulations by \citealt{larwood96}) the diffusive equations adopted
here might not be adequate. To resolve this issue we clearly need
detailed numerical simulations of warp propagation in thin and viscous
discs.

\section*{Acknowledgements}

We thank Andrew King, Steve Lubow, Gordon Ogilvie, Martin Rees and
Marta Volonteri for inspiring discussions and Cathie Clarke for a
careful reading of the manuscript.

\bibliographystyle{mn2e} 
\bibliography{lodato}

\end{document}